\documentclass[aps,prd,floatfix,superscriptaddress,onecolumn,preprintnumbers]{revtex4} \usepackage{amssymb}
\usepackage{amsmath}
\usepackage{amsfonts}
\usepackage{epsfig}             
\usepackage{graphicx}
\usepackage{stackrel}
\usepackage{tabularx}
\usepackage{color}
\usepackage{dsfont}
\usepackage[T1]{fontenc}
\definecolor{darkerblue}{rgb}{0.0,0.0,0.5}
\newcommand{\seq}{\begin{subequations}}
	\newcommand{\sen}{\end{subequations}}

\newcommand{\eq}{\begin{eqnarray}}
\newcommand{\en}{\end{eqnarray}}

\def\nn{\nonumber}

\begin{document}
\preprint{RBI-ThPhys-2022-37}
	
	\title{Dark SU(2) Stueckelberg portal} 
	
	\author{Valery~E.~Lyubovitskij} 
	\affiliation{Institut f\"ur Theoretische Physik, Universit\"at T\"ubingen, \\
		Kepler Center for Astro and Particle Physics, \\ 
		Auf der Morgenstelle 14, D-72076 T\"ubingen, Germany} 
	\affiliation{Departamento de F\'\i sica y Centro Cient\'\i fico
		Tecnol\'ogico de Valpara\'\i so-CCTVal, \\ 
		Universidad T\'ecnica Federico Santa Mar\'\i a, Casilla 110-V, Valpara\'\i so, Chile}
	\affiliation{Millennium Institute for Subatomic Physics at
		the High-Energy Frontier (SAPHIR) of ANID, \\
		Fern\'andez Concha 700, Santiago, Chile}
	\author{Alexey~S.~Zhevlakov} 
	\affiliation{Bogoliubov Laboratory of Theoretical Physics, JINR, 141980 Dubna, Russia} 
	\affiliation{Matrosov Institute for System Dynamics and 
		Control Theory SB RAS, \\  Lermontov str., 134, 664033, Irkutsk, Russia } 
	\author{Aliaksei~Kachanovich} 
	\affiliation{Rudjer Bo$\breve{s}$kovi$\acute{c}$ Institute, Division of Theoretical Physics, 
		Bijeni$\acute{c}$ka 54, HR-10000 Zagreb, Croatia} 
	\author{Serguei~Kuleshov} 
	\affiliation{Millennium Institute for Subatomic Physics at
		the High-Energy Frontier (SAPHIR) of ANID, \\
		Fern\'andez Concha 700, Santiago, Chile}
		\affiliation{Departamento de Ciencias F\'isicas, 
		Universidad Andres Bello, Sazi\'e 2212, Piso 7, Santiago, Chile}
	
	\date{\today}
	
	\begin{abstract}
		
    	        We study the non-Abelian $SU(2)_D$ extension of the 
		$U(1)_D$ Stueckelberg portal, which plays the role of 
		mediator between the Standard Model (SM) and dark sector (DS). 
		This portal is specified by the Stueckelberg mechanism for 
		generation of dark gauge boson masses. 
		The proposed $U(1)_D \otimes SU(2)_D$ Stueckelberg portal has 
		a connection with SM matter fields, in analogy with the familon model. 
		We derive bounds on the couplings of dark portal bosons and SM particles,
                which govern diagonal and nondiagonal flavor 
		transitions of quarks and leptons. 
		
	\end{abstract}
		
	\maketitle
	
	\section{Introduction}
	
	The Standard Model (SM) of particle physics is a
	unified gauge theory of strong and electroweak interactions, 
	which allows one to perform a precise description and explanation 
	of most of data extracted at worldwide facilities wherein, 
	 there are signals of new physics, which cannot be 
	explained within the SM and, therefore, require its extensions. 
        These include, in particular, the muon anomalous magnetic moment
        with last measurements at Fermilab~\cite{Muong-2:2021ojo},
        where the current deviation of the SM prediction and experimental data 
        lies at the $\sim 4.2 \sigma$ level~\cite{Aoyama:2020ynm}. 
	Besides this, there are further unresolved puzzles,
        such as the strong $CP$ problem and rare meson 
	decays~\cite{Pospelov:2017kep,Zhevlakov:2020bvr,Zhevlakov:2019ymi,%
	Zhevlakov:2018rwo,Gutsche:2016jap}, 
	flavor nonuniversality~\cite{Ema:2016ops,Buras:2021btx,Crivellin:2021sff},
        the $b-s$ quark anomaly~\cite{Aebischer:2019mlg},
        neutrino mass generation, and etc. These puzzles initiated many efforts
        for SM extensions and new physics searches. 
	
        The existence of dark matter (DM) is required by a wide spectrum of 
        gravitational, astrophysical, and cosmological phenomena.
        DM significantly contributes to the mass of the Universe.
        However, its precise nature is not yet known. 
	The search for DM created an idea of portals between SM and DM particles.  
	In this respect the main popular portals are Higgs~\cite{Arcadi:2019lka},
        axion~\cite{Nomura:2008ru}, axion-like  particles~\cite{Bauer:2017ris,%
        Bauer:2018uxu,Bauer:2019gfk,Zhevlakov:2022vio,Kaneta:2016wvf},
        vector~\cite{Fortuna:2020wwx,Buras:2021btx}, and sterile
        neutrino~\cite{Escudero:2016tzx} ones. Recently, we proposed the $U(1)_D$
        dark gauge invariant vector portal between particles of the SM and
        Dark Sector (DS)~\cite{Kachanovich:2021eqa}, where the
	of mass of dark photon $A'$ is generated via the Stueckelberg mechanism
        (see also Refs.~\cite{Kribs:2022gri,Ruegg:2003ps}).  
	Additional scalar field $\sigma$ occurring in that approach 
	is an unphysical degree of freedom, which plays the role of ghost field. 
	There is a popular idea of interaction of gauge field and SM fermions based on
        the ansatz of a familon (or flavons)
        in the literature~\cite{Feng:1997tn,Bauer:2019gfk,Cornella:2019uxs}. 
	Such an interaction mechanism between dark photon and SM fermions leads 
	to a rich phenomenology including novel information on couplings preserving and
        violating lepton symmetries, e.g., lepton flavor violation (LFV).  
    
    Phenomenological studies of the dark photon~\cite{Holdom:1985ag} 
    have been performed using different scenarios and particle 
    content (see, e.g., Refs.~\cite{Buras:2021btx,Fayet:2016nyc}). 
    In particular, the dark photon could interact not only with a QED 
    photon-induced so-called kinetic mixing term~\cite{Holdom:1985ag},
    but also with leptonic pair including neutrinos (so-called $Z'$ boson). 
    Additional vector gauge bosons have been extensively studied 
    and searched during three decades~\cite{Altarelli:1989ff}. 
    Especially, very promising studies have been 
    performed at CERN, Fermilab, and other experimental facilities 
    \cite{CMS:2016ifc,ATLAS:2017eqx,ATLAS:2019erb,ATLAS:2020uiq,%
    NA64:2016oww,NA64:2017vtt,NA64:2018lsq,NA64:2021acr,%
    NA64:2022rme,Andreev:2022hxz}. One should stress 
    that phenomenological models considered before were not 
    limited by simple kinetic mixing between SM and new gauge
    bosons~\cite{Pankov:2019yzr,Osland:2022ryb,Osland:2020onj,%
    Fuks:2017vtl}. Indeed, 
    other possible scenarios have been also considered. 
    	
    The main motivation of present paper is an extension of the Abelian
    $U(1)_D$ Stueckelberg portal~\cite{Kachanovich:2021eqa} by adding
    the non-Abelian $SU(2)_D$ sector because it gives an opportunity
    to study the processes with charged dark gauge bosons and currents.  
    In particular, inclusion of additional particles from the $SU(2)_D$  
    portal could give a chance to understand existing deviations 
    between SM and experiments. In particular, we introduce the $SU(2)_D$
    triplet of dark gauge bosons (DGBs). Such an extension is natural and has 
    the additional benefit of opening a window for couplings between SM 
    and DS particles via charged currents. 
    Note that the Abelian $U(1)_D$ Stueckelberg portal is based on
    the coupling of a dark photon with neutral current formed by SM fermions,
    while inclusion of the $SU(2)_D$ DGBs interacting with both charged 
    and neutral currents can be performed in analogy to the weak sector 
    of SM. Thereby, all properties of the Stueckelberg portal such as LFV
    are implemented. Besides, the LFV effect~\cite{Heeck:2016xwg}, 
    such $SU(2)_D$ extension induces lepton number violation.
   	    	
    Discussion of the dark fermion sector, which is closely related to 
    charge dark gauge bosons $W^{\prime\pm}$, is beyond the scope 
    of present paper. By the way, we note that it is a very interesting topic
    (see, e.g., recent discussion in Ref.~\cite{Belyaev:2022shr}).  
    Instead, we focus on phenomenological aspects of the $SU(2)_D$ vector  
    Stueckelberg portal and derivation of bounds on its couplings  
    with SM fermions based on current deviations between the SM 
    and data. In our study, we intend to consider both LFV and non-LFV
    lepton decays as a tool for such analysis. We also  mention that
    the new $W'$ bosons have been intensively studied inn different theoretical
    approaches beyond the SM. In particular, the $W's$ bosons have been proposed
    by several theoretical approaches based on gauge extensions
    of the Standard Model~\cite{Altarelli:1989ff,Eichten:1984eu},
    extradimensional~\cite{Randall:1999ee,Randall:1999vf,Davoudiasl:2000wi},
    technicolor~\cite{Lane:2002sm,Eichten:2007sx},
    and composite Higgs models ~\cite{Agashe:2004rs,Giudice:2007fh} models. 
    Experimental searches of the $W'$ bosons have been performed at LHC by the ATLAS
    and CMS Collaborations~\cite{ATLAS:2019isd,ATLAS:2019lsy,CMS:2018iye,CMS:2022yjm}
    based on the production of the $W'$ in proton-proton
    collisions, which led to the constraint of the $W'$ mass in the 
    range 0.15-6 TeV. In our approach we assume that the dark $W'$ boson masses
    do not exceed TeV region.  
	
    The paper is organized as follows.
    In Sec.~\ref{Nonabelian_DS}, we define the Lagrangian for the non-Abelian extension
    of the dark sector. In Sec.~\ref{Wpbounds}, we estimate bounds on the couplings
    of the charged DGBs imposed by various muon decay processes, with
    $\mu \to e \nu_i \bar{\nu}_j$ in Sec.~\ref{muenunu} and $\mu \to e \gamma$
    in Sec.~\ref{muegamma}. The latter is relevant for the muon $g-2$ anomaly.
    Finally, in Sec.~\ref{conclusions} we present our conclusions. 
		
	\section{Non-Abelian extension of Dark Sector}
	\label{Nonabelian_DS} 
	
	In this section, we discuss an extension of the Abelian dark 
	Stueckelberg portal to the non-Abelian case. In particular, 
	we propose the existence of three additional DGBs and 
	three additional DSBs associated with $SU(2)_D$ group, 
	which are required by gauge invariant principle.
        Our formalism is based of effective SM+DS Lagrangian ${\cal L}_{\rm SM+DS}$,
        which after spontaneous breaking of electroweak symmetry to electromagnetic
        group $U(1)_Y \otimes SU(2)_L \to U(1)_{\rm em}$
        is manifestly gauge-invariant
        under the product of electromagnetic group and groups of DS and 
        $U(1)_{\rm em} \otimes U(1)_D \otimes SU(2)_D$. 
        First we specify the fields of our Lagrangian.
        For convenience, in this paper we use the notations introduced before for
        the Abelian case~\cite{Kachanovich:2021eqa} and introduce additional ones
        relevant for the non-Abelian dark sector sector. 
        The SM sector contains 
        fundamental fermions --- left $q_L^{im}$ and right ($U_R^i$, $D_R^i$) quarks, 
	left and right leptons $(\ell_L^{im}, \ell_R^i)$ fields), gauge fields 
        (weak gauge bosons $W^\pm$, $Z^0$ and photon $A$)
        and scalar Higgs field $H$. Left doublets and right 
	singlets of quarks and leptons are specified as
	$q_L^{1m} = (u_L,d_L)$,
        $q_L^{2m} = (c_L,s_L)$,
        $q_L^{3m} = (t_L,b_L)$, 
	$U_R^1 = u_R$,
        $U_R^2 = c_R$,
        $U_R^3 = t_R$, 
	$D_R^1 = u_R$,
        $D_R^2 = c_R$,
        $D_R^3 = b_R$,  
	$L^{1m} = (\nu_{eL},e_L)$,
        $L^{2m} = (\nu_{\mu L},\mu_L)$, 
	$L^{3m} = (\nu_{\tau L},\tau_L)$,
        $R^1 = e_R$,
        $R^2 = \mu_R$, and
	$R^3 = \tau_R$. The indices  
        $i,j = 1,2,3$ number the fermion generations,
        while $m,n$ 
        denote the $SU(2)$ weak isospin and dark 
        indices, respectively.  
        
        The DS sector contains singlet dark fermion  
        $\chi$,
        gauge bosons (singlet $A'$ and triplet ${\bf W'}$),and
        scalars (singlet $\sigma$ and triplet ${\bf S}$).
        The matrices ${\bf S}$ and ${\bf W}'$ are specified as 
	\eq 
	{\bf S} = 
	\left(
	\begin{array}{cc}
		S^0/\sqrt{2} &   S^+ \\[1mm]
		S^-          & - S^0/\sqrt{2} \\[1mm]
	\end{array} 
	\right) \,, \quad 
	{\bf W}^{'} = 
	\left(
	\begin{array}{cc}
		W^{'0}/\sqrt{2} &   W^{'+}          \\[1mm]
		W^{'-}          & - W^{'0}/\sqrt{2} \\[1mm]  
	\end{array} 
	\right) \,.
	\en
        Triplets of gauge bosons $W'$ and scalars $S$ have electric charge.
        In particular, $W^{'\pm}$ and $S^\pm$ have charges $Q= \pm 1$, respectively,
        while $W^{'0}$, $S^0$, and dark fermions are electrically
        neutral. 
        
        Now we define the covariant derivatives acting on fermions and scalars.
        In the case of the SM sector the change will be in adding of the terms containing
        dark gauge fields in the covariant derivatives acting on the left
        and right fermions 
        \eq
	\Big(iD^L\Big)_\mu^{mn}  \to \Big(iD^L\Big)_\mu^{mn}
        - \frac{g_{W'}}{2} \, {\bf W}^{' mn}_\mu 
        - g_{A'} \, \delta^{mn}  \, A'_\mu \,, 
	\qquad 
	\Big(iD^R\Big)_\mu \to \Big(iD^R\Big)_\mu - g_{A'} \, A'_\mu  \,,
        \en 
        where $g_{W'}$ and $g_{A'}$ are the gauge couplings associated with
        $SU(2)_D$ and $U(1)_D$ groups, respectively, In the case of the DS, the covariant derivates
        acting on scalar fields $(D_\mu\sigma)$ and $(D_\mu{\bf S})$
        and fermions are defined respectively, as
        \eq 
	(D_\mu\sigma)           = \partial_\mu {\bf \sigma} - M_{A'}  \, A'_\mu\,, \qquad 
	(D_\mu{\bf S})^{mn}      = \partial_\mu {\bf S}^{mn} - M_{W'}  \, {\bf W}^{' mn}_\mu
	\en
        and 
        \eq
	\Big(iD_{\chi}\Big)_\mu = i \, \partial_\mu 
	- g_{A'}  \, A'_\mu \,. 
	\en 
        Note that covariant derivatives acting on dark scalars 
        $(D_\mu\sigma)$ and $(D_\mu{\bf S})$  
	include dark gauge bosons (singlet $A'$)  
	and triplet ($W^{'\pm}, W^{'0}$) having finite masses $M_{A'}$ and 
	$M_{W'}$, respectively. 
	The covariant derivatives ($D_\mu\sigma$) and ($D_\mu{\bf S}$) therefore contain 
	DS gauge boson masses, generated via the Stueckelberg 
	mechanism~\cite{Stueckelberg:1938zz,Ruegg:2003ps}.   
	The $U(1)_D$ gauge boson $A'$ is called the dark photon and it has 
	the mass $M_{A'}$. The scalar Stueckelberg fields $\sigma$ and $\bf{S}$ 
	play the role of supplementary Goldstone bosons generating masses
        of dark photon and dark triplet gauge bosons ${\bf W}'$.
	Such an idea was considered before for the new $Z^\prime$ boson in 
	Ref.~\cite{Bell:2016uhg} (see also discussion in Ref.~\cite{Gherghetta:2019coi}).  
        DGBs acquire the masses in a manifestly gauge-invariant form. 
	Finite masses of scalars violate $U(1)_D \otimes SU(2)_D$ 
	symmetry (for a review see Ref.~\cite{Ruegg:2003ps}).  
	Extension of the Stueckelberg mechanism on non-Abelian field leads
        to known problems of renormalizability and unitarity. 
	Critical remarks and efforts to resolve these problems have 
	been discussed in detail in Ref.~\cite{Ruegg:2003ps}. Note that, due 
	to the scale of new physics $\Lambda$ is much larger than the scale of SM 
	($\Lambda \gg \Lambda_{\rm SM})$; thus, we do not go to higher loops and restrict 
	ourselves to one-loop approximation. Therefore, the problems
        mentioned above are not so critical for our purposes.
	Note that, for convenience we write all terms in Lagrangian involving 
	the ${\bf S}$ triplet in a simplified form, which follows from 
	Ref.~\cite{Delbourgo:1986wz}, where this field was derived using 
	adjoint representation of the $SU(2)_D$ gauge group         
	\eq 
	U(x) = \exp\biggl[- ig_{W'} \frac{{\bf S}(x)}{2M_{W'}} \biggr] \,. 
	\en 
	In this approach, 
	\eq 
	\partial_\mu{\bf S}(x) = \frac{2 M_{W'}}{g_{W'}} 
	\, i\partial_\mu U(x) \, U^{-1}(x) \,. 
	\en
        We propose that the Stueckelberg  mechanism~\cite{Stueckelberg:1938zz,Ruegg:2003ps} 
	for generating masses of gauge fields is extended to the group 
	$U(1)_D \otimes SU(2)_D$.  

        The stress tensors for dark gauge bosons are defined as 
	\eq 
	A_{\mu\nu}' = 
	\partial_\mu A_{\nu}' - 
	\partial_\nu A_{\mu}' \,, \qquad 
	{\bf W}_{\mu\nu}' =  
	\partial_\mu{\bf W}_{\nu}' - 
	\partial_\nu{\bf W}_{\mu}' + 
	\frac{i g_{W'}}{2} 
	\, [{\bf W}_{\mu}',{\bf W}_{\nu}'] \,. 
	\en 
	The  $U(1)_D \otimes SU(2)_D$  
        gauge transformations of dark fermions, 
	scalars, and gauge bosons are specified as  
	\eq 
	& &
	A^\prime_\mu \to A^\prime_\mu 
	+ \frac{i}{g_{A'}} \, \partial_\mu\Omega_{A'} 
	\, \Omega^{-1}_{A'} 
	\,, \qquad \hspace*{2cm}
	A_{\mu\nu}' \to A_{\mu\nu}' \,, \nonumber\\
	& &
	{\bf W}^\prime_\mu \to 
	\Omega_{W'} \, {\bf W}^\prime_\mu \, 
	\Omega^{-1}_{W'} + \frac{2i}{g_{W'}} \, 
	\partial_\mu\Omega_{W'} \, \Omega^{-1}_{W'} 
	\,, \qquad 
	{\bf W}_{\mu\nu}' \to \Omega_{W'} \, {\bf W}_{\mu\nu}' \, 
	\Omega^{-1}_{W'} \,, \
	\nonumber\\
	& &\partial_\mu\sigma \to \partial_\mu\sigma 
	+ \frac{i M_{A'}}{g_{A'}} \, \partial_\mu\Omega_{A'} \, 
	\Omega^{-1}_{A'} 
	\,, \qquad \hspace{1.4cm} (D_\mu {\bf \sigma} )
        \to  (D_\mu {\bf \sigma} )
	\,, \nonumber\\
	& &U(x) \to \Omega_{W'}(x) \, U(x) \,, 
	\qquad \qquad\qquad\qquad\qquad
	\partial_\mu{\bf S} \to
        \Omega_{W'} \,  \partial_\mu{\bf S} \, \Omega_{W'}^{-1} \,+\, 
	\frac{2i M_{W'}}{g_{W'}} \, \partial_\mu\Omega_{W'} \, 
	\Omega^{-1}_{W'}\,, \\
	& &  (D_\mu {\bf S})  
	\to \Omega_{W'} \, (D_\mu {\bf S}) \, \Omega_{W'}^{-1}\,,
        \qquad\qquad\qquad\qquad
	{\rm Tr}\Big[ (D_\mu {\bf S}) (D_\mu {\bf S}) \Big] \to 
	{\rm Tr}\Big[ (D_\mu {\bf S}) (D_\mu {\bf S}) \Big] \,, \nonumber\\
	& &\chi \to \Omega_{A'} \, \chi \,,
        \hspace*{5.2cm} 
	i\not\!\! D_{\chi} \chi \to \Omega_{A'}  
	\, i\not\!\! D_{\chi} \, \chi 
        \,, \nonumber
	\en 
	where 
	\eq
	\Omega_{W'}(x) = \exp\Big[ \frac{i}{2} \vec{\mathbf\theta}_{W'}(x)
        \vec{\mathbf\tau} \Big] \,, 
	\qquad 
	\Omega_{A'}(x) = \exp\Big[ i {\theta}_{A'}(x) \Big] 
	\en 
	are the matrices of the fundamental $SU(2)_D$ 
        and $U(1)_D$ transformations.

        Now we specify effective Lagrangian of our approach
        ${\cal L}_{\rm SM+DS}$ combining SM and DS: 
        \eq 
	{\cal L}_{\rm SM+DS} = {\cal L}_{\rm SM}
        + {\cal L}_{\rm DS} 
	+ \Delta{\cal L} \,. 
	\en
        This Lagrangian is by construction 
	a low-energy Lagrangian, which is an extrapolation 
	of new physics Lagrangian including the DS sector to the SM scale 
	$\Lambda_{\rm SM} \simeq M_{W^\pm/Z^0} \simeq 100$ GeV. 
        Here ${\cal L}_{\rm SM}$ is the term describing dynamics of the SM sector
        including the coupling of the SM fermions with DS gauge fields via extension
        of covariant derivatives introduced above, 
        ${\cal L}_{\rm DS}$ is the term describing dynamics of the DS
        including the coupling of the dark fermions with SM gauge fields
        via extension of covariant derivatives introduced above
        and two terms describing the coupling of SM and DS - 
	GB mixing term ${\cal L}_{\rm mix}$ and a term describing additional 
	coupling of DSBs with SM fields $\Delta{\cal L}$  allowing a violation
        of lepton flavor and number violation and 
	violation of the Glashow-Illiopulos-Maiani (GIM) mechanism~\cite{GIM}
	in the quark sector.

        Next we specify the terms ${\cal L}_{\rm DS}$ and ${\cal L}_{\rm int}$.
        The DS Lagrangian ${\cal L}_{\rm DS}$ is given by: 
        \eq  
        {\cal L}_{\rm DS} &=& - 
	\frac{1}{4} \, {A}_{\mu\nu}' {A}^{\prime \mu\nu} \,-\,
	\frac{1}{4} \, {\bf W}_{\mu\nu}' {\bf W}^{\prime \mu\nu} \,+\,
	\frac{1}{2} \, (D_\mu\sigma) (D^\mu\sigma) 
	+ 
	\frac{1}{2} \, {\rm Tr}\Big[ (D_\mu {\bf S}) (D^\mu {\bf S}) \Big] 
	+  \bar\chi \, (i\not\!\!D_\chi - m_\chi) \, \chi 
        \nonumber\\
        &-& \frac{1}{2\xi_{W'}} \text{Tr} 
	\left( \partial_{\mu} {\bf W}'^\mu  
	+ \xi_{W'} M_{W'} {\bf S} \right)^2 - \frac{1}{2\xi_{A'}} 
	\left( \partial_{\mu} {A}'^\mu  
	+ \xi_{A'} M_{A'} {\bf \sigma} \right)^2 
	\,. 
	\en  
	One should stress that quantization of the dark gauge field is
        required to add the gauge-fixing term into the dark $U(1)_D$ and
        $SU(2)_D$ sectors~\cite{Kors:2005uz} (last two terms
        in ${\cal L}_{\rm DS}$), 
	where $\xi_{W'(A')}$ is an arbitrary gauge parameter corresponding 
	to the $W'(A')$ gauge boson.
        It provides ``decoupling'' of gauge bosons and 
	corresponding scalars particles with vanishing mixed terms:  
	\eq 
	\mathcal{L}_{DS} &=&  
	-     \frac{1}{4} \, {A}_{\mu\nu}' {A}^{\prime \mu\nu} 
	\,-\, \frac{1}{4} \, {\bf W}_{\mu\nu}' {\bf W}^{\prime \mu\nu} 
	\,+\, \frac{1}{2} \, \text{Tr} \, (\partial_\mu {\bf S} \partial^\mu {\bf S}) 
	\,+\, \frac{M_{W'}^{2}}{2} \, \text{Tr} ({\bf W_{\mu}} {\bf W^{\mu}}) 
	\,+\, \frac{M_{A'}^2}{2} A'_\mu A'^\mu\nonumber\\
	&-&   \frac{1}{2 \xi_{W'}}\text{Tr}\,(\partial_\mu {\bf W}' \partial^\mu {\bf W}')
	\,-\, \frac{\xi_{W'}}{2} M_{W'}^2 \text{Tr}\left({\bf S}^2\right)  
	\,+\, \frac{1}{2}\, \partial_\mu \sigma \partial^\mu \sigma 
	\,-\, \frac{1}{2 \xi_{A'}} (\partial_{\mu} A'^{\mu})^2 
	\,-\, \frac{\xi_{A'}}{2} M_{A'}^2 \sigma^2 \nonumber\\
	&+&  \bar\chi_L \, (i\not\!\!D_\chi^L - m_\chi) \, 
	\chi_L \,+\,  \bar\chi_R \, (i\not\!\! D_\chi^R - m_\chi) \, \chi_R \,. 
	\en
        We note that the masses of the $\sigma$ and triplet $\bf{S}$ are proportional
        to the gauge parameter $\xi_i$, signaling that these fields 
        are unphysical. 
        In the gauge, we are using, the dark boson propagator takes the form 
        \eq\label{eq:dark_propagator-1}
        D^{\mu\nu}(k; \xi_i) = \frac{1}{k^2 - M^2} \, 
        \biggl[ g^{\mu\nu} - \frac{k^\mu k^\nu \, (1-\xi_i)}{k^2  - \xi_i M^2} 
        \biggr] \,.
        \en 
        where $M$ is the mass of the dark gauge boson. 

        The interaction $\Delta{\cal L}$ Lagrangian is constructed 
	by analogy with the familon model proposed in Ref.~\cite{Feng:1997tn} 
	for the hypothetical familon field. The scalar fields are unphysical 
	and are to be switched off by the choice of gauge fixing: 
	\eq 
	\Delta{\cal L} &=& \frac{1}{\Lambda} \, 
	(D_\mu {\bf \sigma}) \,  \sum_{ij} 
	\biggl[ 
	\bar q_L^{i} c^{ij}_{\sigma} \gamma^\mu q_L^{j}    \,+\, 
	\bar U_R^i c^{ij}_{U \sigma} \gamma^\mu U_R^j        \,+\, 
	\bar D_R^i c^{ij}_{D \sigma} \gamma^\mu D_R^j        \,+\, 
	\bar L^{i} d^{ij}_{L \sigma} \gamma^\mu L^{j}      \,+\, 
	\bar R^i d^{ij}_{R \sigma} \gamma^\mu R^j 
	\biggr] \nonumber\\
	&+& \frac{1}{\Lambda} \, (D_\mu {\bf S})^{mn} \, \sum_{ij} 
	\biggl[ 
	\bar q_L^{im} c^{ij}_{\bf S} \gamma^\mu q_L^{jn} \,+\, 
	\bar L^{im} d^{ij}_{\bf S}   \gamma^\mu L^{jn} 
	\biggr]
        \,, 
	\label{eq:Stuckelberg-portal-1}
	\en
	where $\Lambda$ is the scale of new physics. 
	Here, $c^{ij}$ and $d^{ij}$ are the $3 \times 3$ hermitian matrices 
	containing the couplings of dark scalars with the SM fermions, and include 
	effects of lepton flavor and number violation, and 
	violation of the Glashow-Illiopulos-Maiani (GIM) mechanism~\cite{GIM}
	in the quark sector. The parameter $\Lambda$ is the characteristic scale
        of this effective operator, defining when it opens up in terms of
        renormalizable interactions of an UV completion. 	
	
	As was pointed out in Ref.~\cite{Bauer:2017ris} 
	after spontaneous breaking of electroweak symmetry in the SM,
        one should diagonalize the fermion mass matrices by means of
        unitary transformations
	\eq
	Y_U = (V_L^q)^\dagger \, y_U \, W_R^U\,, \quad 
	Y_D = (V_L^q)^\dagger \, y_D \, W_R^D\,, \quad  
	Y_\ell = (V_L^\ell)^\dagger \, y_\ell \, W_R^\ell\,.
	\en 
	Here $V_L^q$ and $V_L^\ell$ are the transformation matrices acting 
	on the left-handed quarks and leptons, respectively, and 
	$W_R^U$, $W_R^D$, and $W_R^\ell$ are the transformation 
	matrices acting on right singlets,  and 
        $y^{ij}$ are the $3 \otimes 3$ Yukawa matrices of 
	couplings between two scalar doublets and SM fermions
        before spontaneous breaking of electroweak symmetry. 
        The matrices $V_L^q$, $V_L^\ell$ $W_R^U$, and $W_R^D$  
        rotate the couplings of dark scalars with the SM fermions as 
	\eq 
	c_{\sigma  }  &\to& C_{\sigma}   = (V_L^q)^\dagger \, c_{\sigma}   \, V_L^q\,, 
	\nonumber\\
	c_{U \sigma}  &\to& C_{U \sigma} = (W_R^U)^\dagger \, c_{U \sigma} \, W_R^U\,, 
	\nonumber\\
	c_{D \sigma}  &\to& C_{D \sigma} = (W_R^D)^\dagger \, c_{D \sigma} \, W_R^D\,, 
	\nonumber\\
	d_{L \sigma}  &\to& D_{L \sigma} = (V_L^q)^\dagger \, c_{L \sigma} \, V_L^q\,, \\
	d_{R \sigma}  &\to& D_{R \sigma} = (W_R^D)^\dagger \, d_{R \sigma} \, W_R^U\,, 
	\nonumber\\
	c_{{\bf S}}   &\to& C_{{\bf S}}  = (V_L^q)^\dagger \, c_{{\bf S}}   \, V_L^q\,, 
	\nonumber\\
	d_{{\bf S}}   &\to& D_{{\bf S}}  = (V_L^\ell)^\dagger \, d_{{\bf S}}   \, V_L^\ell\,.
	\nonumber 
	\en 
	
	Note that the resulting coupling of the SM fermions with DS fields is contributed by
        three terms via minimal substitution of the covariant derivatives acting on the SM fermions
        and via additional effective Lagrangian~(\ref{eq:Stuckelberg-portal-1}). The first term 
        does not mix the SM generations, does not violate certain symmetries
        (like lepton flavor and lepton number), and preserves the GIM mechanism: 
        \eq\label{Lint1} 
	{\cal L}_{\rm int; 1} &=& g_{A'}  \, \sum_{i}  \bar{\psi}_i \, \gamma^{\mu} 
	\, A'_\mu \, \psi_i  
	+ \frac{g_{W'}}{2} \, \sum_{i}  \, 
	\bar{\psi}_i^{m} \, \gamma^{\mu} \, {\bf W'}_\mu^{mn} \,
	\left(1 - \gamma_5 \right) \psi_i^{n} \,,
	\en

        After the substitution of the covariant derivatives
        $(\partial \sigma)$ and $(\partial \bf{S})$, we find that the gauge-invariant  
	operator~(\ref{eq:Stuckelberg-portal-1}) additionally generates dimension-4 interactions
        of the dark photon and $\bf{W'}$ bosons 
	with the SM fermions $\psi$, in the form
        \eq\label{Lint2} 
	{\cal L}_{\rm int; 2} &=&
        \sum_{ij}  \bar{\psi}_i \, \gamma^{\mu} 
	\, A'_\mu \, \left(g^{V}_{ij} 
	+ g^{A}_{ij}\gamma_5 \right) \psi_j  
	+ \sum_{ij} g^{V\!A}_{ij} 
	\bar{\psi}_i^{m} \, \gamma^{\mu} \, {\bf W'}_\mu^{mn} \,
	\left(1 - \gamma_5 \right) \psi_j^{n} \,,
	\en
 	where vector $g^{V}$ and axial-vector $g^{A}$ dimensionless 
	couplings are defined as 
	\eq\label{eq:g-V-AV}
	g_{ij}^{V}= \frac{m_{A'}} {\Lambda} v_{ij} \,, & \qquad\qquad
	&  g_{ij}^{A}=\frac{m_{A'}} {\Lambda} a_{ij} \,,
        \\
	v_{ij}=\frac{1}{2}\left(D^{ij}_{R \sigma} + D^{ij}_{L \sigma}\right)_{ij} \,,
        & \qquad\qquad& a_{ij}=\frac{1}{2}\left(D^{ij}_{R \sigma} 
	- D^{ij}_{L \sigma}\right)_{ij}
	\,. 
	\en
	Dimensionless couplings of $W'$ with leptons 
	have the simple form $g^{V\!A}_{ij}=D_{\bf S}^{ij} M_{W^\prime}/\Lambda$. 
	Scalar nonphysical fields can be switched off from consideration 
	in the case of unitary gauge ($\xi\to\infty$), which corresponds 
	to the limit of infinitely large masses of scalars or vanishing 
	of their contribution to physical processes.
        Therefore, the full interaction Lagrangian of
        the SM fermions with DGBs reads
        \eq\label{Lintfull} 
        {\cal L}_{\rm int} = {\cal L}_{\rm int; 1} + {\cal L}_{\rm int; 2}
        =
        \sum_{ij}  \bar{\psi}_i \, \gamma^{\mu} 
	\, A'_\mu \, \left( G^{V}_{ij} 
	+ G^{A}_{ij}\gamma_5 \right) \psi_j  
	+ \sum_{ij} G^{V\!A}_{ij} 
	\bar{\psi}_i^m \, \gamma^{\mu} \, {\bf W'}_\mu^{mn} \,
	\left(1 - \gamma_5 \right) \psi_j^n \,,
	\en
	where
        \eq
        G^{V}_{ij} = g^{V}_{ij}
        + \delta_{ij} g_{A'} \,, \quad
        G^{A}_{ij} \equiv g^{A}_{ij} \,, \quad
        G^{V\!A}_{ij} = g^{V\!A}_{ij}
        + \delta_{ij}  \frac{g_{W'}}{2}  \,.
        \en        

	\section{Bounds on $W^\prime$ couplings with Standard model particles}
	\label{Wpbounds}
	
	In this section, we discuss opportunities to estimate bounds on the couplings 
	of the charged dark gauge boson $W'$ with the SM particles. 
	On the one hand, we base on data extracted from precise measurements,
        and on the other hand we can involve rare decays in our analysis.
        First, we estimate diagonal couplings $G^{V\!A}_{ii}$, 
	which give an additional contribution to the SM processes. 
	Hereinafter, we will concentrate on nondiagonal couplings $G^{V\!A}_{ij}$ ($i\neq j$),
        which can be responsible for a contribution to LFV processes. 
	
	\subsection{Dominant and LFV $\mu \to e \nu_i \bar{\nu}_j$ decays}
	\label{muenunu}
	
	To derive a limit for the $G^{V\!A}_{ii}$ coupling we use 
	one of the most precise  measurements in particle physics --- 
	decay rate $\mu\to e \nu_\mu \bar{\nu}_e$ process. 
	This decay gives a very accurate determination of 
	the Fermi constant $G^\mu_F$: 
	\eq 
	G^{\mu}_F = 1.1663787(6) \times 10^{-5} \text{GeV}^{-2}
        \en 
	at the level of $0.5$ ppm~\cite{MuLan:2010shf,MuLan:2012sih,Gorringe:2015cma}.  
	
	It is clear that the diagonal coupling of dark gauge boson $W'$ 
	gives an additional contribution to the $\mu \to e \nu_\mu \bar\nu_e$ decay rate.
	In our benchmark scenario, we make a conjecture that there is an ambiguity
        for the precise determination of the Fermi constant $G_F$.
        In particular, this constant can be constrained using different data (see, 
	e.g., Ref.~\cite{Crivellin:2021njn}). In particular, $G_F$ can be determined from
        analysis of kaon or $\tau$ decays using unitarity of the Cabibbo-Kobayashi-Maskawa matrix.
        Alternatively, one can use the global electroweak (EW) fit. 
	Obtained values have a small deviation from the value of the $G_F$ extracted
        from weak muon decay. Indeed, such uncertainty is a gap, which constrains possible
        contribution of the $W'$ to the $\mu\to e \nu_\mu \bar{\nu}_e$ muon decay and provides
        bounds on its diagonal or nondiagonal couplings in dependence on
        a type of dark vector boson. 
	
	For calculation of decay widths, we use the well-known formula 
	\eq
	d\Gamma = \frac{4\pi^2}{2 m} |M|^2 d\Phi_n \,, 
	\en
	where $m$ is the mass of decaying particle, $|M|^2$ is the square 
	of amplitude of a process, which defines its dynamics, and $d\Phi_n$ is
        an element of $n$-body phase space. 
	
	The muon decay rate $\mu\to e \nu_\mu \bar{\nu}_e$ in the SM is 
	determined by 
	\eq
	\Gamma(\mu^- \to \nu_\mu e^-\bar{\nu}_e) =\left( \frac{\sqrt{2}g_{EW}^2}{8m_W^2} \right)^2 
	\frac{m_\mu^5}{192\pi^3}(1+\Delta q)  = 
	\frac{(G_F^\mu)^2m_\mu^5}{192\pi^3}(1+\Delta q) \,, 
	\en
	where the quantity $\Delta q$ includes the phase space, QED, 
	and hadronic radioactive corrections (see Ref.~\cite{Crivellin:2021njn}).  
	
        The square of the $\mu\to e \nu_\mu \bar{\nu}_e$ decay amplitude, taking into account of
        the mixing of the SM $W$ and dark $W'$ bosons, is given by 
\begin{equation} 
     |M|^2=-\frac{(s+t) \left(s+t-m_{\mu }^2\right) \left(g_{EW}^2 \left(t-m_{W'}^2\right)
    +(G^{V\!A})^2 \left(t-m_W^2\right)\right)^2}{ \left(m_{W'}^2-t\right)^2 \left(t-m_W^2\right)^2} \,, 
\end{equation}
where the Mandelstam kinematical variables are defined as 
$s=m_\mu^2-2(p_\mu p_e)$, $t=m_\mu^2-2(p_\mu p_{\nu_1})$, and
$u=m_\mu^2-2(p_\mu p_{\nu_2})$~\cite{ParticleDataGroup:2020ssz}, and
$g_{EW}$ is electroweak coupling. 
We obtain an expression for the decay width as a function of parameters of
the new vector boson (mass of $m_{W'}$ and couplings with leptons $G_{ij}^{V\!A}$). 

\begin{figure}[h]
		\includegraphics[width=0.6\textwidth, trim={0cm 0cm 0cm 0cm},clip]{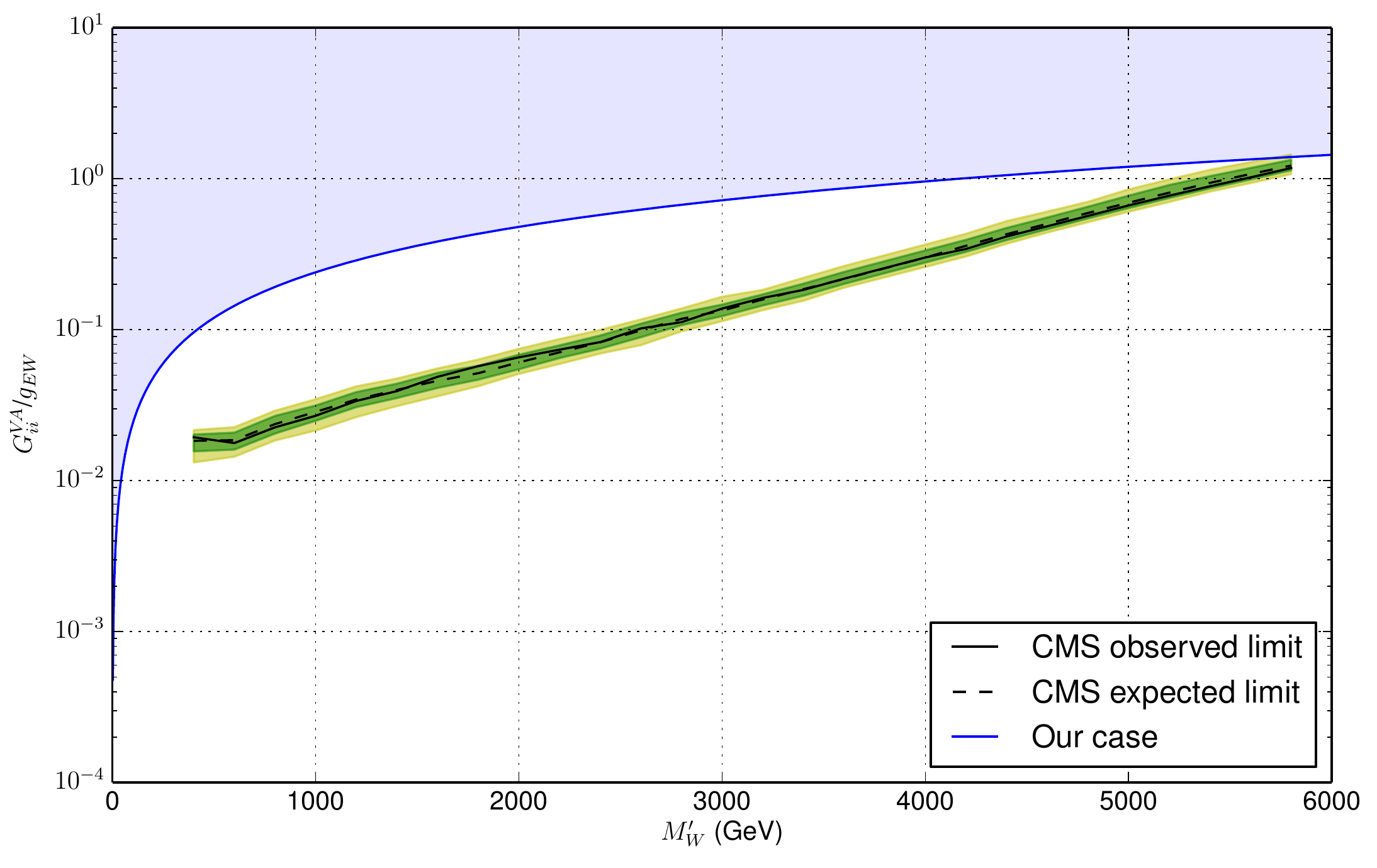}
	\caption{Bounds on the ratio $G^{V\!A}_{ij}/g_{EW}$ from possible contribution to the $\mu \to e \nu\nu$ 
	decay rate (dirty blue line) as a function of $m_{W^\prime}$ 
	in the range 2 GeV $-$ 6 TeV. Black solid and dashed lines are CMS observed and expected limit, respectively, with one standard deviation and two standard deviations (green and yellow areas) \cite{CMS:2022yjm}. These two limits can be compared only in the universality case of diagonal coupling dark boson with quarks and leptons.}
	\label{muenunu_mWprime}
\end{figure}

Using the difference between data of the global
EW fit presented in Ref.~\cite{Crivellin:2021njn} 
$G^{EW}_F|_{full}=1.16716(39) \times 10^{-5} \,\, \text{GeV}^{-2}$ 
and $G^\mu_F$ from $\mu\to e \nu_\mu \bar{\nu}_e$  decay, 
we can derive limits for the couplings of $W'$ with 
the SM particles using the muon width as input data parameter.  
Bounds on the coupling ratio $G^{V\!A}_{ij}/g_{EW}$ as function of the $W'$ mass are shown 
in Fig.~\ref{muenunu_mWprime}. These bounds in the case of the universality scenario
of the coupling of the dark $W'$ boson with quarks and leptons can be compared with
limits extracted by the CMS Collaboration \cite{CMS:2022yjm} (see Fig.~\ref{muenunu_mWprime}). In this scenario at $m_{W'} \ge 0.5$ TeV, the CMS limits are more stronger 
and our bounds established from study of the famous muon decay complement in this
range of the dark boson mass. In the case of the nonuniversal coupling of dark boson
$W'$ with quark and leptons (general scenarios) 
our limits deduced from the $\mu\to e \nu_\mu \bar{\nu}^c_e$ decay cannot
be directly compared with the CMS constraints. 
The difference between values of the $G_F$ constant defined from
different experimental data gives strong limits on diagonal couplings of
the dark charge boson $W^{'\pm}$  with the SM fermions. 

We also want to mention about a landscape of study a possible dark charge $W'$ boson.
The existed investigation is based on studies of different
process \cite{ATLAS:2023vxg,CMS:2022yjm} and different
models \cite{Lindner:2016bgg,Osland:2020onj,Osland:2022ryb}.
As output of such studies, different bounds on the masses, 
couplings, and other parameters of additional $W'$ and $Z'$ bosons
have been proposed in the literature. Sometimes different bounds from different
studies cannot be compared directly.
Here, we would like to stress that novel $W'$ and $Z'$ bosons
have been proposed in literature before in different
extensions of SM at TeV scales, including
extradimensional models, technicolor, and composite Higgs
(for review see, e.g., \cite{Langacker:2008yv}).
Also the existence of these bosons has been searched at the LHC.
Some constraints of the $W'$ and $Z'$ bosons masses in the TeV region
have been performed in dependence of their coupling strengths with
the SM fermions. On the other hand, searches of light $W'$ and $Z'$
are also attracted a lot of interest for resolving existing
puzzles and anomalies. 
In particular, recently the ATLAS Collaboration \cite{ATLAS:2023vxg}
set upper limits on the $Z'$
production cross section times the decay branching fraction
of the $p p \to Z' \mu^+ \mu^- \mu^+ \mu^-$ process, varying
from 0.31 to 4.3 fb at 95\% C.L.,
in a $Z'$ mass range 5-81 GeV, from which the coupling strength of the $Z'$
to muons above 0.003 to 0.2 (depending on the $Z'$ mass) are excluded in
the same mass range.

Here we suppose scenarios that the $W^\prime$
is heavier than the $\tau$ lepton. It is necessary to forbid invisible decays
of leptons into neutrinos and dark bosons, which can further decay into light dark fermions. 
On the other hand, $W^\prime$ should be heavier than the SM $W^\pm/Z$ bosons, as otherwise
one could see the $W^\prime$ pair creation in the final state of physical reactions, e.g., 
in the $e^+ e^-$ annihilation or hadron-hadron collisions. 
Masses of dark fermions interacting with dark charge vector current 
should be larger than masses of the SM leptons. 
We see that the $U(1)_D$ portal can be connected with sub-GeV particles, whereas  
dark mediators from the $SU(2)_D$ sector are weak interactive massive particles,  
which have larger masses in comparison with one of the SM weak bosons. 

In the universality scenario of interaction $W'$ with all types of the SM leptons, 
we have bound on $G^{V\!A}_{ii}$ less than $ 10^{-1} \div 3 \times 10^{-4}$   
in the following range of the $W'$ mass 2 GeV $\div$ 1 TeV.   
Such limits correspond to the $G^{V\!A}_{ij}$ nondiagonal coupling in the case 
when the corresponding decay is induced by exchange of intermediate neutral dark 
boson $W'^0$. In the case of a neutral boson we have less constraints as 
in case of the charge partners $W'^\pm$. Therefore, the range of the 
neutral dark boson mass in Fig.~\ref{muenunu_mWprime} is extended to a region 
of smaller values of mass. Moreover, we will further show that these bounds are 
the strongest ones that can be obtained from pure
LFV decay  $\mu^- \to e^-\bar{\nu}_\mu\nu_e$. 
Two LFV processes will be considered later in our manuscript. 
 
	\begin{figure}[t]
		\includegraphics[width=0.6\textwidth, trim={0cm 0cm 0cm 0cm},clip]{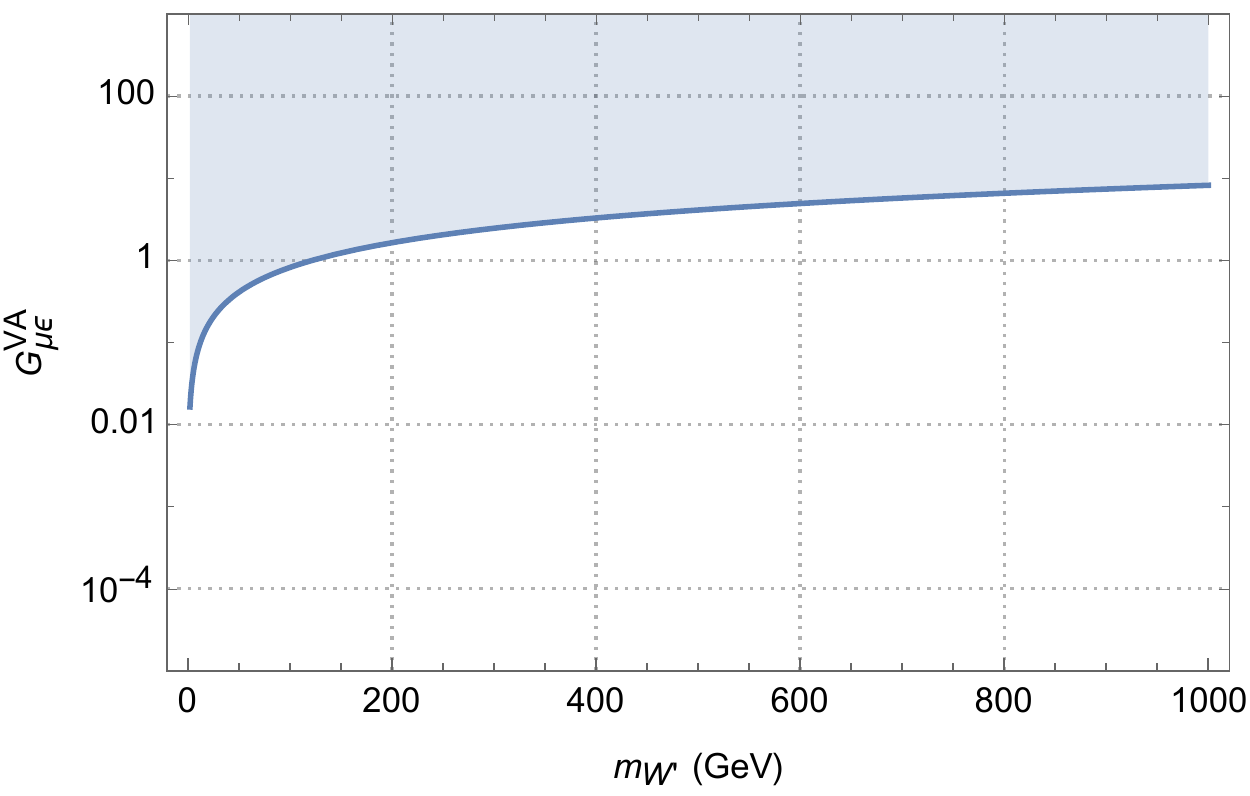}
	\caption{Bounds on the $\mu-e$ non-diagonal lepton flavor coupling 
	from existing LFV data 
	$\mu^-\to \nu_e\bar{\nu}_\mu e^-$ \cite{Freedman:1993kz}. 
	The shaded area is a closed band for dark boson couplings.}
	\label{g_mu_e_Wprime}
\end{figure} 

Because of the structure of our interaction~(\ref{Lintfull}), 
we establish the same limits for the nondiagonal couplings of both neutral and 
charged dark bosons with leptons with specific flavor and mass. 
In case of nondiagonal couplings, we do not have restrictions on the mass of neutral boson
from charge conservation as it occurs for charged dark bosons.  
It is different from a general scenario with a dark $Z'$ boson, 
where we can describe 
interaction with leptons and neutrino independently: we consider neutral $W'^0$ component
in couple with charged $W^{\prime \pm}$ bosons. In our approach 
dark photon $A'$ plays the role of neutral dark boson $Z'$~\cite{Kachanovich:2021eqa}. 
Therefore, due to possible interference of $A'$ and $W'^0$ one can 
establish new limits on couplings with leptons and neutrinos, which could 
be different from those derived in Ref.~\cite{Kachanovich:2021eqa}.  

Next we can calculate decay width of the $W'$ 
to $l\bar{\nu}_l$ pair by analogy to the similar decay in SM: 
\eq
\Gamma(W' \to e^-\bar{\nu}_e) \simeq \frac{(G^{V\!A}_{e\bar{\nu}_e})^2 m_{W'}}{48\pi} 
\en
and as example for $m_{W^\prime}=200$ GeV we have 
$\Gamma(W' \to e^-\bar{\nu}_e) =6.5 \times 10^{-4}$ MeV.  
It is by $\sim 6$ orders less than we have for weak $W$ boson decay width.  

Nondiagonal interaction couplings of the $W'$ dark boson can be constrained from LFV processes,
e.g., from muon LFV decay $\mu^- \to e^-\bar{\nu}_\mu\nu_e$.  
The best upper limit on the branching of this decay is 1.2\%~\cite{Freedman:1993kz}. 
We  note that this existing limit on the $\mu^- \to e^-\bar{\nu}_\mu\nu_e$ decay rate 
is not very sensitive and feeble competitive with many other experiments which 
limit most of new physics scenarios. Contrariwise we stress that this LFV decay 
gives limitation directly on one of the nondiagonal components, i.e., on coupling $G^{V\!A}_{\mu e}$. 
Bounds on this coupling are presented in Fig.~\ref{g_mu_e_Wprime}.  Because the interaction couplings in our model for charged and neutral $W'$ are universal,
we see that famous weak muon decay $\mu\to e \nu_\mu \bar{\nu}_e$
gives more stronger limits on the coupling constants in comparison with limits derived from LFV decay,
e.g., from the $\mu^- \to e^-\bar{\nu}_\mu\nu_e$ decay mode. 

\subsection{LFV muon decay $\mu\to e\gamma$}
\label{muegamma}
\begin{figure}[b]

		\includegraphics[width=0.3\textwidth, 
		trim={2.5cm 22.5cm 10cm 2cm},clip]{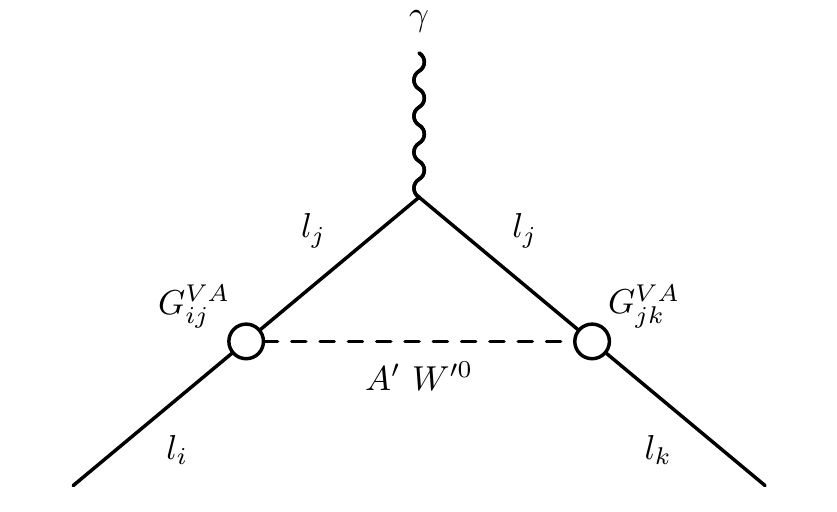}

	\caption{Feynman diagrams with contribution of 
	dark photon $A'$ and charge dark boson $W'$ 
	to the LFV process $l_i \to l_k\gamma$.}
	\label{fig:DecayLFV}
\end{figure} 

New dark vector bosons can potentially explain LFV effects. In this respect, 
$\mu \to e \gamma$ decay is a good laboratory for the study of such effects. 
Feynman diagrams contributing to this process taking into account 
$A'$ and $W'^{\pm}$ bosons are presented in Fig.~\ref{fig:DecayLFV}. 
Note that the diagram with a pair of intermediate charged bosons $W'^{\pm}$ 
has an additional suppression factor in comparison with the dark photon $A'$ exchange. 

Existing experimental limits for the branchings of the LFV 
lepton decays $l_i \to l_k \gamma$ are~\cite{ParticleDataGroup:2020ssz}  
\eq
\mathrm{Br}(\mu\to e\gamma ) &<& 4.2 \times 10^{-13}\,, \nonumber\\
\mathrm{Br}(\tau\to e\gamma ) &<& 3.3 \times 10^{-8}\,, \nonumber\\
\mathrm{Br}(\tau\to \mu\gamma ) &<& 4.2 \times 10^{-8} \,.
\en
The general matrix element of this LFV process can be parametrized as
\eq
iM_{ik} = ie \, \epsilon^{\mu}(q) \, \bar{u}_k(p_2,m_k) \, \left[ \frac{i}{2m_i}\sigma_{\mu\nu}q^{\nu} 
F_M +\frac{i}{2 m_i}\sigma_{\mu\nu}q^{\nu}\gamma_5 F_D \right] \, u_i(p_1,m_i) \,, 
\en
where the square of matrix element is
\eq
|M_{ik}|^2 &= & m_i^2 \, \biggl[1-\frac{m_k^2}{m_i^2}\biggr]^2 \, 
\biggl(|F_M|^2+ |F_D|^2\biggr) 
\en
and the decay width is given by
\eq
\Gamma(l_i \to l_k\gamma) &=& \frac{1}{2m_i} \int \frac{d^3p_2 
\, d^3q}{4E_2 E_q  \, (2\pi)^6} \, 
(2\pi)^4 \, \delta^{(4)}(p_1-p_2-q) \ |M_{ik}|^2  
\\
&=& \frac{\alpha}{2} \, m_i \, \biggl(|F_M|^2+ |F_D|^2 \biggr) \,. \nn
\en
Here, $F_M$ and $F_D$ are the magnetic and dipole form factors, respectively, and 
$\alpha =e^2/(4\pi) = 1/137.036$ is the fine-structure constant.  
More explicitly, when the $A'$ propagates in the loop (right-hand diagram
in Fig.~\ref{fig:DecayLFV}), for a general $l_i \to l_k \gamma$ with an $i$- or $k$- lepton
in the loop, we obtain
\eq
\label{F1}
F_M &=& \frac{1}{16\pi^2} \left[\left(G^{V}_{ik}G^{V}_{ii}
  + G^{A}_{ik}G^{A}_{ii}\right) 
  h_2^{V}(x_\mu)+\left(G^{V}_{ik}G^{V}_{kk}
  + G^{A}_{ik}G^{A}_{kk}\right)h_3^{V}(x_\mu)\right]\,, \nn\\
F_D &=& \frac{1}{16\pi^2} \left[\left(G^{V}_{ik}G^{A}_{ii}
  + G^{A}_{ik}G^{V}_{ii}\right) 
  h_2^{V}(x_\mu)+\left(G^{V}_{ik}G^{A}_{kk}
  + G^{A}_{ik}G^{V}_{kk}\right)h_3^{V}(x_\mu)\right]
\,,
\en
whereas for $\mu \to e \gamma$, with the $\tau$ lepton propagating 
in the loop and with double LFV coupling, we have:
\eq
F_M &=&  \frac{1}{16\pi^2} \left(\frac{m_\mu}{m_\tau}\right)
\left[ G^{V}_{\mu \tau} 
  G^{V}_{e \tau}h_1^{V}(x_\tau) -
  G^{A}_{\mu \tau}G^{A}_{e \tau}h_1^{V}(x_\tau)\right] \nn
\,, 
\nonumber\\
F_D &=& \frac{1}{16\pi^2} \left(\frac{m_\mu}{m_\tau}\right)\left[ G^{V}_{\mu \tau} 
G^{A}_{e \tau}h_1^{V}(x_\tau)- G^{A}_{\mu \tau}G^{V}_{e \tau}h_1^{V}(x_\tau)\right] 
\label{F2}
\,,
\en
where  $x_i = m_{A'}^2/m_i^2$. Expressions for the loop functions 
$h_i^{V}(x_i)$ in the approximation $m_e \ll m_\mu \ll m_\tau$ 
are shown in the Appendix~\ref{App_Loops}.

	\begin{figure}[t]
	  \includegraphics[width=0.6\textwidth, trim={0cm 0cm 0cm 0cm},clip]
                          {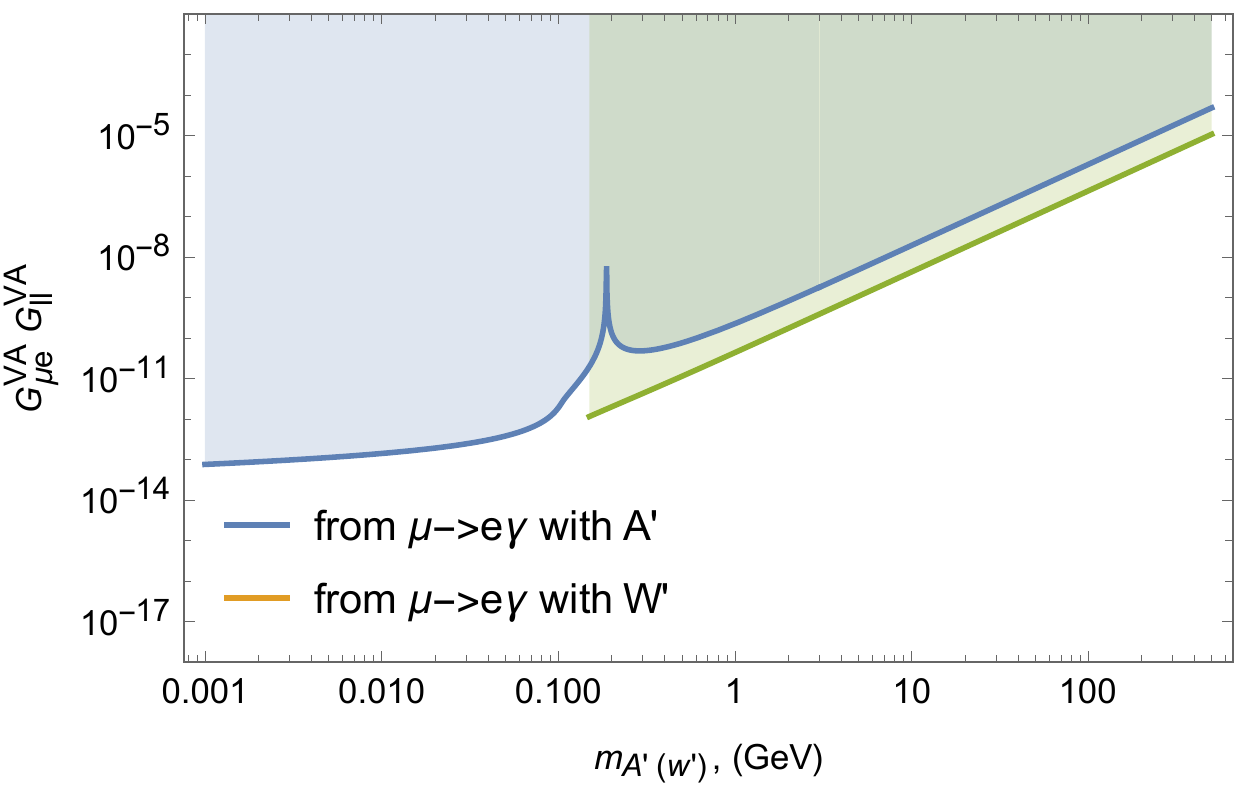}
	                  \caption{Bounds to production $\mu-e$ nondiagonal
                            lepton flavor coupling
        and diagonal coupling from existing 
	FV data $\mu^-\to e^- \gamma$~\cite{ParticleDataGroup:2020ssz}.
        The shaded area is a closed band for dark boson couplings
        (blue for dark photon and only $G^V$ contribution and green for $W'^0$
        neutral dark boson due to full contribution from vector and
        axial-vector parts).}
	\label{g_mu_eLFVdecay}
\end{figure} 

The matrix element corresponding to the loop LFV diagram 
induced by the coupling of a neutral $W'$ boson with a photon  
and contributing to the $l_i \to l_k \gamma$ process 
is specified by two form factors 
\eq
\label{F3}
F_M &=& \frac{1}{16\pi^2} \left[\left(G^{VA}_{ik}G^{VA}_{ii}\right) 
h^{W^0}(x_\mu)+2\left(G^{VA}_{ik}G^{VA}_{kk}\right)h_3^{V}(x_\mu)\right]\,, \nn\\
F_D &=& \frac{1}{16\pi^2} \left[\left(G^{VA}_{ik}G^{VA}_{ii}\right) 
h^{W^0}(x_\mu)+2\left(G^{VA}_{ik}G^{VA}_{kk}\right)h_3^{V}(x_\mu)\right]\,,
\en
where $x=m_{W'^0}^2/m_l^2$. Double LFV contributions are 
neglected here. 

Dependence on the masses of dark bosons $A'$ and $W'$
is presented in Fig.~\ref{g_mu_eLFVdecay}. 
The peaks in Fig.~\ref{g_mu_eLFVdecay} are connected with the behavior of 
the loop integrals $h_i(x)$ near the point $x=1$ located in the vicinity of
the vector boson production threshold. 
To resolve this problem, one needs to include in the analysis 
the finite width $\Gamma_{A'} \sim \tau_{A'}^{-1} \sim G_{ij}^2$ 
of the dark vector boson in the Breit-Wigner propagator. 
The latter is dominated by the decay width of the $A'$ to the leptonic pair. 

We make conservative estimate for coupling production 
$G^{V\!A}_{ll} G^{V\!A}_{e\mu}$ in proposition that diagonal
couplings $G^{V\!A}_{ll}$ are equal to each other.  
Corresponding bounds are shown in Fig.~\ref{g_mu_eLFVdecay}. 
It means that bounds on 
$G^{V\!A}_{ll} G ^{V\!A}_{e\mu}$ from
$\mu\to e\gamma$ LFV decay for neutral vector bosons are 
strict for a light mass boson.  
Limits on the $A'$ couplings correspond to the fact that the bounds for the 
neutral dark boson $W'^0$ are divided by a factor of two, because they have 
the same mechanism of interaction with SM matter 
governed by both vector and axial couplings. Further discussion of bounds
on couplings $A'$ dark photon has been done in Ref.~\cite{Kachanovich:2021eqa}
(here, we note that in Ref. \cite{Kachanovich:2021eqa} in the plot
of the curve for limit on the LFV coupling derived from the
decay $\mu \to e \gamma$ the factor proportional to
$10^{-6}$ was lost). With taking into account with factor the limits
on the LFV couplings of the dark gauge bosons with SM fermions
are consistent in this paper and Ref.~\cite{Kachanovich:2021eqa}). 
At the same time, it is important to stress a constantly increasing research
interest to study sub-GeV 
dark candidates which are the main goal in running and planning experiments
for searching dark matter at fixed target regime such as NA64 SPS
at CERN \cite{NA64:2022rme,NA64:2022yly,NA64:2021acr,NA64:2020xxh,%
NA64:2019auh,Banerjee:2019pds,NA64:2018lsq,NA64:2017vtt,NA64:2016oww}, 
M$^3$ \cite{Kahn:2018cqs}, and LDMX \cite{Mans:2017vej}. LFV coupling between dark
boson and SM particles can explain existing experimental anomalies, wherein different models provide different limits on mass and couplings
of the $Z'$ boson \cite{Profumo:2013sca,Kelso:2013zfa,Alves:2015pea}.

\section{conclusion}
\label{conclusions}

We have proposed a phenomenological Lagrangian approach
that combines the SM and DM sectors based on the Stueckelberg mechanism
for generation of masses of 
dark $U(1)_D$ and $SU(2)_D$ gauge bosons due to the presence of 
dark scalar Stueckelberg fields. These scalar fields play 
the role of Goldstone bosons.
A novel part consists in adding of non-Abelian part
to the dark vector $U(1)_D$ portal. 
We derived bounds on diagonal and nondiagonal interaction couplings
between $SU(2)_D$ dark gauge bosons and the SM leptons. 
In particular, we established limits on the couplings using data from
canonical weak muon decay  $\mu\to e \nu_\mu \bar{\nu}_e$ supposing 
that some correction to its decay rate is possible and given the ambiguity in
the definition of value of $G_F$. Additionally, we have used the phenomenology
of lepton flavor-violating processes to derive limits on the $W'$ couplings. 
In this paper, we concentrated on thee gauge boson sector of our approach. 
It would be interesting to extend our formalism on the
dark fermion sector and consider applications on other rare lepton decays. 

\begin{acknowledgments} 
	
        This work was funded by BMBF (Germany) ``Verbundprojekt 05P2021 (ErUM-FSP T01) -
	Run 3 von ALICE am LHC: Perturbative Berechnungen von Wirkungsquerschnitten
	f\"ur ALICE'' (F\"orderkennzeichen: 05P21VTCAA), by ANID PIA/APOYO AFB180002
        and AFB220004 (Chile), by FONDECYT (Chile) under Grant No. 1191103
	and by ANID$-$Millen\-nium Program$-$ICN2019\_044 (Chile). 
		The work of A.S.Z. is supported by grant of AYSS JINR (22-302-02). 
	The work was supported by the Foundation for the Advancement of Theoretical 
        Physics and Mathematics "BASIS".
	
\end{acknowledgments}	

\appendix
\section{Loop functions which is used at calculation $\mu \to e \gamma$ LFV decay}
\label{App_Loops}

In this appendix, we present analytical expressions for the loop integrals occurring
in the amplitude of the LFV decays $l_i \to l_k \gamma$ 
for different new particle channels and leptons propagating in the loop,
used in Eqs.~(\ref{F1})-(\ref{F3}).
All results for the form factors have been numerically and analytically 
cross-checked using the Mathematica packages Package-X~\cite{Patel:2015tea},  
FeynHelpers~\cite{Shtabovenko:2016whf} and 
FeynCalc \cite{Shtabovenko:2020gxv,Shtabovenko:2016sxi,Mertig:1990an}.

For the dark heavy neutral boson $W'$, the loop integrals read
\eq
h^{W'^0}(x) &=& 8\left( \text{Li}_2\left(1-x\right)- \text{Li}_2\left(\frac{2}{2-x+\sqrt{(x-4) x}}\right)
+ \text{Li}_2\left(\frac{2}{x+\sqrt{(x-4) x}}\right)\right)
 +2 x (2 x^2-9 x+9) \frac{\log(x)}{x-1} \nn\\
 &+&2 (-4 x+\frac{1}{x}+8)+4 \log^2\left(\frac{x+\sqrt{(x-4) x}-2)}{x+\sqrt{(x-4) x}}\right)
 -4 \sqrt{(x-4) x} (2 x-3) \log\left(\frac{x+\sqrt{(x-4) x}}{2 \sqrt{x}}\right) \,.
\en
For the $A'$ dark boson, the loop integrals are given by 
\eq
h^{V}_1(x) 
&=& -\frac{(4x^3-3x^2-6x^2\ln(x)-1)}{x(1-x)^3} \,, 
\\
h^V_2(x) &=&   
2 \left(2 \text{Li}_2(1-x)-2 \text{Li}_2\left(\frac{2}{-x+\sqrt{(x-4) x}+2}\right)+2
\text{Li}_2\left(\frac{2}{x+\sqrt{(x-4) x}}\right)-2 x +2 \log(x)\right.
\\&+& \left. \log
^2\left(\frac{x + \sqrt{x - 4) x }}{2 x }\right)+\frac{(x+1) ((x-4) x+2) \log
	\left(x\right)}{x-1}-2 x \sqrt{(x-4) x} 
\log \left(\frac{\sqrt{x}+\sqrt{(x-4) }}{2}\right)+1\right) \,, \nn
\\
h^{V}_3(x) &=&-4 x+4 (x-1)^2  \ln \left(\frac{x}{x-1}\right)+6 \,.
\en

\bibliography{biblSU2}

\begin{thebibliography}{85}
\expandafter\ifx\csname natexlab\endcsname\relax\def\natexlab#1{#1}\fi
\expandafter\ifx\csname bibnamefont\endcsname\relax
  \def\bibnamefont#1{#1}\fi
\expandafter\ifx\csname bibfnamefont\endcsname\relax
  \def\bibfnamefont#1{#1}\fi
\expandafter\ifx\csname citenamefont\endcsname\relax
  \def\citenamefont#1{#1}\fi
\expandafter\ifx\csname url\endcsname\relax
  \def\url#1{\texttt{#1}}\fi
\expandafter\ifx\csname urlprefix\endcsname\relax\def\urlprefix{URL }\fi
\providecommand{\bibinfo}[2]{#2}
\providecommand{\eprint}[2][]{\url{#2}}

\bibitem[{\citenamefont{Abi et~al.}(2021)}]{Muong-2:2021ojo}
\bibinfo{author}{\bibfnamefont{B.}~\bibnamefont{Abi}} \bibnamefont{et~al.}
  (\bibinfo{collaboration}{Muon g-2 Collaboration}), \bibinfo{journal}{Phys.
  Rev. Lett.} \textbf{\bibinfo{volume}{126}}, \bibinfo{pages}{141801}
  (\bibinfo{year}{2021}), \eprint{2104.03281}.

\bibitem[{\citenamefont{Aoyama et~al.}(2020)}]{Aoyama:2020ynm}
\bibinfo{author}{\bibfnamefont{T.}~\bibnamefont{Aoyama}} \bibnamefont{et~al.},
  \bibinfo{journal}{Phys. Rept.} \textbf{\bibinfo{volume}{887}},
  \bibinfo{pages}{1} (\bibinfo{year}{2020}), \eprint{2006.04822}.

\bibitem[{\citenamefont{Pospelov and Tsai}(2018)}]{Pospelov:2017kep}
\bibinfo{author}{\bibfnamefont{M.}~\bibnamefont{Pospelov}} \bibnamefont{and}
  \bibinfo{author}{\bibfnamefont{Y.-D.} \bibnamefont{Tsai}},
  \bibinfo{journal}{Phys. Lett. B} \textbf{\bibinfo{volume}{785}},
  \bibinfo{pages}{288} (\bibinfo{year}{2018}), \eprint{1706.00424}.

\bibitem[{\citenamefont{Zhevlakov and Lyubovitskij}(2020)}]{Zhevlakov:2020bvr}
\bibinfo{author}{\bibfnamefont{A.~S.} \bibnamefont{Zhevlakov}}
  \bibnamefont{and} \bibinfo{author}{\bibfnamefont{V.~E.}
  \bibnamefont{Lyubovitskij}}, \bibinfo{journal}{Phys. Rev. D}
  \textbf{\bibinfo{volume}{101}}, \bibinfo{pages}{115041}
  (\bibinfo{year}{2020}), \eprint{2003.12217}.

\bibitem[{\citenamefont{Zhevlakov
  et~al.}(2019{\natexlab{a}})\citenamefont{Zhevlakov, Gutsche, and
  Lyubovitskij}}]{Zhevlakov:2019ymi}
\bibinfo{author}{\bibfnamefont{A.~S.} \bibnamefont{Zhevlakov}},
  \bibinfo{author}{\bibfnamefont{T.}~\bibnamefont{Gutsche}}, \bibnamefont{and}
  \bibinfo{author}{\bibfnamefont{V.~E.} \bibnamefont{Lyubovitskij}},
  \bibinfo{journal}{Phys. Rev. D} \textbf{\bibinfo{volume}{99}},
  \bibinfo{pages}{115004} (\bibinfo{year}{2019}{\natexlab{a}}),
  \eprint{1904.08154}.

\bibitem[{\citenamefont{Zhevlakov
  et~al.}(2019{\natexlab{b}})\citenamefont{Zhevlakov, Gorchtein, Hiller~Blin,
  Gutsche, and Lyubovitskij}}]{Zhevlakov:2018rwo}
\bibinfo{author}{\bibfnamefont{A.~S.} \bibnamefont{Zhevlakov}},
  \bibinfo{author}{\bibfnamefont{M.}~\bibnamefont{Gorchtein}},
  \bibinfo{author}{\bibfnamefont{A.~N.} \bibnamefont{Hiller~Blin}},
  \bibinfo{author}{\bibfnamefont{T.}~\bibnamefont{Gutsche}}, \bibnamefont{and}
  \bibinfo{author}{\bibfnamefont{V.~E.} \bibnamefont{Lyubovitskij}},
  \bibinfo{journal}{Phys. Rev. D} \textbf{\bibinfo{volume}{99}},
  \bibinfo{pages}{031703} (\bibinfo{year}{2019}{\natexlab{b}}),
  \eprint{1812.00171}.

\bibitem[{\citenamefont{Gutsche et~al.}(2017)\citenamefont{Gutsche,
  Hiller~Blin, Kovalenko, Kuleshov, Lyubovitskij, Vicente~Vacas, and
  Zhevlakov}}]{Gutsche:2016jap}
\bibinfo{author}{\bibfnamefont{T.}~\bibnamefont{Gutsche}},
  \bibinfo{author}{\bibfnamefont{A.~N.} \bibnamefont{Hiller~Blin}},
  \bibinfo{author}{\bibfnamefont{S.}~\bibnamefont{Kovalenko}},
  \bibinfo{author}{\bibfnamefont{S.}~\bibnamefont{Kuleshov}},
  \bibinfo{author}{\bibfnamefont{V.~E.} \bibnamefont{Lyubovitskij}},
  \bibinfo{author}{\bibfnamefont{M.~J.} \bibnamefont{Vicente~Vacas}},
  \bibnamefont{and}
  \bibinfo{author}{\bibfnamefont{A.}~\bibnamefont{Zhevlakov}},
  \bibinfo{journal}{Phys. Rev. D} \textbf{\bibinfo{volume}{95}},
  \bibinfo{pages}{036022} (\bibinfo{year}{2017}), \eprint{1612.02276}.

\bibitem[{\citenamefont{Ema et~al.}(2017)\citenamefont{Ema, Hamaguchi, Moroi,
  and Nakayama}}]{Ema:2016ops}
\bibinfo{author}{\bibfnamefont{Y.}~\bibnamefont{Ema}},
  \bibinfo{author}{\bibfnamefont{K.}~\bibnamefont{Hamaguchi}},
  \bibinfo{author}{\bibfnamefont{T.}~\bibnamefont{Moroi}}, \bibnamefont{and}
  \bibinfo{author}{\bibfnamefont{K.}~\bibnamefont{Nakayama}},
  \bibinfo{journal}{JHEP} \textbf{\bibinfo{volume}{01}}, \bibinfo{pages}{096}
  (\bibinfo{year}{2017}), \eprint{1612.05492}.

\bibitem[{\citenamefont{Buras et~al.}(2021)\citenamefont{Buras, Crivellin,
  Kirk, Manzari, and Montull}}]{Buras:2021btx}
\bibinfo{author}{\bibfnamefont{A.~J.} \bibnamefont{Buras}},
  \bibinfo{author}{\bibfnamefont{A.}~\bibnamefont{Crivellin}},
  \bibinfo{author}{\bibfnamefont{F.}~\bibnamefont{Kirk}},
  \bibinfo{author}{\bibfnamefont{C.~A.} \bibnamefont{Manzari}},
  \bibnamefont{and} \bibinfo{author}{\bibfnamefont{M.}~\bibnamefont{Montull}},
  \bibinfo{journal}{JHEP} \textbf{\bibinfo{volume}{06}}, \bibinfo{pages}{068}
  (\bibinfo{year}{2021}), \eprint{2104.07680}.

\bibitem[{\citenamefont{Crivellin and Hoferichter}(2021)}]{Crivellin:2021sff}
\bibinfo{author}{\bibfnamefont{A.}~\bibnamefont{Crivellin}} \bibnamefont{and}
  \bibinfo{author}{\bibfnamefont{M.}~\bibnamefont{Hoferichter}},
  \bibinfo{journal}{Science} \textbf{\bibinfo{volume}{374}},
  \bibinfo{pages}{1051} (\bibinfo{year}{2021}), \eprint{2111.12739}.

\bibitem[{\citenamefont{Aebischer et~al.}(2020)\citenamefont{Aebischer,
  Altmannshofer, Guadagnoli, Reboud, Stangl, and Straub}}]{Aebischer:2019mlg}
\bibinfo{author}{\bibfnamefont{J.}~\bibnamefont{Aebischer}},
  \bibinfo{author}{\bibfnamefont{W.}~\bibnamefont{Altmannshofer}},
  \bibinfo{author}{\bibfnamefont{D.}~\bibnamefont{Guadagnoli}},
  \bibinfo{author}{\bibfnamefont{M.}~\bibnamefont{Reboud}},
  \bibinfo{author}{\bibfnamefont{P.}~\bibnamefont{Stangl}}, \bibnamefont{and}
  \bibinfo{author}{\bibfnamefont{D.~M.} \bibnamefont{Straub}},
  \bibinfo{journal}{Eur. Phys. J. C} \textbf{\bibinfo{volume}{80}},
  \bibinfo{pages}{252} (\bibinfo{year}{2020}), \eprint{1903.10434}.

\bibitem[{\citenamefont{Arcadi et~al.}(2020)\citenamefont{Arcadi, Djouadi, and
  Raidal}}]{Arcadi:2019lka}
\bibinfo{author}{\bibfnamefont{G.}~\bibnamefont{Arcadi}},
  \bibinfo{author}{\bibfnamefont{A.}~\bibnamefont{Djouadi}}, \bibnamefont{and}
  \bibinfo{author}{\bibfnamefont{M.}~\bibnamefont{Raidal}},
  \bibinfo{journal}{Phys. Rept.} \textbf{\bibinfo{volume}{842}},
  \bibinfo{pages}{1} (\bibinfo{year}{2020}), \eprint{1903.03616}.

\bibitem[{\citenamefont{Nomura and Thaler}(2009)}]{Nomura:2008ru}
\bibinfo{author}{\bibfnamefont{Y.}~\bibnamefont{Nomura}} \bibnamefont{and}
  \bibinfo{author}{\bibfnamefont{J.}~\bibnamefont{Thaler}},
  \bibinfo{journal}{Phys. Rev. D} \textbf{\bibinfo{volume}{79}},
  \bibinfo{pages}{075008} (\bibinfo{year}{2009}), \eprint{0810.5397}.

\bibitem[{\citenamefont{Bauer et~al.}(2017)\citenamefont{Bauer, Neubert, and
  Thamm}}]{Bauer:2017ris}
\bibinfo{author}{\bibfnamefont{M.}~\bibnamefont{Bauer}},
  \bibinfo{author}{\bibfnamefont{M.}~\bibnamefont{Neubert}}, \bibnamefont{and}
  \bibinfo{author}{\bibfnamefont{A.}~\bibnamefont{Thamm}},
  \bibinfo{journal}{JHEP} \textbf{\bibinfo{volume}{12}}, \bibinfo{pages}{044}
  (\bibinfo{year}{2017}), \eprint{1708.00443}.

\bibitem[{\citenamefont{Bauer et~al.}(2019)\citenamefont{Bauer, Heiles,
  Neubert, and Thamm}}]{Bauer:2018uxu}
\bibinfo{author}{\bibfnamefont{M.}~\bibnamefont{Bauer}},
  \bibinfo{author}{\bibfnamefont{M.}~\bibnamefont{Heiles}},
  \bibinfo{author}{\bibfnamefont{M.}~\bibnamefont{Neubert}}, \bibnamefont{and}
  \bibinfo{author}{\bibfnamefont{A.}~\bibnamefont{Thamm}},
  \bibinfo{journal}{Eur. Phys. J. C} \textbf{\bibinfo{volume}{79}},
  \bibinfo{pages}{74} (\bibinfo{year}{2019}), \eprint{1808.10323}.

\bibitem[{\citenamefont{Bauer et~al.}(2020)\citenamefont{Bauer, Neubert,
  Renner, Schnubel, and Thamm}}]{Bauer:2019gfk}
\bibinfo{author}{\bibfnamefont{M.}~\bibnamefont{Bauer}},
  \bibinfo{author}{\bibfnamefont{M.}~\bibnamefont{Neubert}},
  \bibinfo{author}{\bibfnamefont{S.}~\bibnamefont{Renner}},
  \bibinfo{author}{\bibfnamefont{M.}~\bibnamefont{Schnubel}}, \bibnamefont{and}
  \bibinfo{author}{\bibfnamefont{A.}~\bibnamefont{Thamm}},
  \bibinfo{journal}{Phys. Rev. Lett.} \textbf{\bibinfo{volume}{124}},
  \bibinfo{pages}{211803} (\bibinfo{year}{2020}), \eprint{1908.00008}.

\bibitem[{\citenamefont{Zhevlakov et~al.}(2022)\citenamefont{Zhevlakov,
  Kirpichnikov, and Lyubovitskij}}]{Zhevlakov:2022vio}
\bibinfo{author}{\bibfnamefont{A.~S.} \bibnamefont{Zhevlakov}},
  \bibinfo{author}{\bibfnamefont{D.~V.} \bibnamefont{Kirpichnikov}},
  \bibnamefont{and} \bibinfo{author}{\bibfnamefont{V.~E.}
  \bibnamefont{Lyubovitskij}}, \bibinfo{journal}{Phys. Rev. D}
  \textbf{\bibinfo{volume}{106}}, \bibinfo{pages}{035018}
  (\bibinfo{year}{2022}), \eprint{2204.09978}.

\bibitem[{\citenamefont{Kaneta et~al.}(2017)\citenamefont{Kaneta, Lee, and
  Yun}}]{Kaneta:2016wvf}
\bibinfo{author}{\bibfnamefont{K.}~\bibnamefont{Kaneta}},
  \bibinfo{author}{\bibfnamefont{H.-S.} \bibnamefont{Lee}}, \bibnamefont{and}
  \bibinfo{author}{\bibfnamefont{S.}~\bibnamefont{Yun}},
  \bibinfo{journal}{Phys. Rev. Lett.} \textbf{\bibinfo{volume}{118}},
  \bibinfo{pages}{101802} (\bibinfo{year}{2017}), \eprint{1611.01466}.

\bibitem[{\citenamefont{Fortuna et~al.}(2021)\citenamefont{Fortuna, Roig, and
  Wudka}}]{Fortuna:2020wwx}
\bibinfo{author}{\bibfnamefont{F.}~\bibnamefont{Fortuna}},
  \bibinfo{author}{\bibfnamefont{P.}~\bibnamefont{Roig}}, \bibnamefont{and}
  \bibinfo{author}{\bibfnamefont{J.}~\bibnamefont{Wudka}},
  \bibinfo{journal}{JHEP} \textbf{\bibinfo{volume}{02}}, \bibinfo{pages}{223}
  (\bibinfo{year}{2021}), \eprint{2008.10609}.

\bibitem[{\citenamefont{Escudero et~al.}(2017)\citenamefont{Escudero, Rius, and
  Sanz}}]{Escudero:2016tzx}
\bibinfo{author}{\bibfnamefont{M.}~\bibnamefont{Escudero}},
  \bibinfo{author}{\bibfnamefont{N.}~\bibnamefont{Rius}}, \bibnamefont{and}
  \bibinfo{author}{\bibfnamefont{V.}~\bibnamefont{Sanz}},
  \bibinfo{journal}{JHEP} \textbf{\bibinfo{volume}{02}}, \bibinfo{pages}{045}
  (\bibinfo{year}{2017}), \eprint{1606.01258}.

\bibitem[{\citenamefont{Kachanovich et~al.}(2022)\citenamefont{Kachanovich,
  Kovalenko, Kuleshov, Lyubovitskij, and Zhevlakov}}]{Kachanovich:2021eqa}
\bibinfo{author}{\bibfnamefont{A.}~\bibnamefont{Kachanovich}},
  \bibinfo{author}{\bibfnamefont{S.}~\bibnamefont{Kovalenko}},
  \bibinfo{author}{\bibfnamefont{S.}~\bibnamefont{Kuleshov}},
  \bibinfo{author}{\bibfnamefont{V.~E.} \bibnamefont{Lyubovitskij}},
  \bibnamefont{and} \bibinfo{author}{\bibfnamefont{A.~S.}
  \bibnamefont{Zhevlakov}}, \bibinfo{journal}{Phys. Rev. D}
  \textbf{\bibinfo{volume}{105}}, \bibinfo{pages}{075004}
  (\bibinfo{year}{2022}), \eprint{2111.12522}.

\bibitem[{\citenamefont{Kribs et~al.}(2022)\citenamefont{Kribs, Lee, and
  Martin}}]{Kribs:2022gri}
\bibinfo{author}{\bibfnamefont{G.~D.} \bibnamefont{Kribs}},
  \bibinfo{author}{\bibfnamefont{G.}~\bibnamefont{Lee}}, \bibnamefont{and}
  \bibinfo{author}{\bibfnamefont{A.}~\bibnamefont{Martin}},
  \bibinfo{journal}{Phys. Rev. D} \textbf{\bibinfo{volume}{106}},
  \bibinfo{pages}{055020} (\bibinfo{year}{2022}), \eprint{2204.01755}.

\bibitem[{\citenamefont{Ruegg and Ruiz-Altaba}(2004)}]{Ruegg:2003ps}
\bibinfo{author}{\bibfnamefont{H.}~\bibnamefont{Ruegg}} \bibnamefont{and}
  \bibinfo{author}{\bibfnamefont{M.}~\bibnamefont{Ruiz-Altaba}},
  \bibinfo{journal}{Int. J. Mod. Phys. A} \textbf{\bibinfo{volume}{19}},
  \bibinfo{pages}{3265} (\bibinfo{year}{2004}), \eprint{hep-th/0304245}.

\bibitem[{\citenamefont{Feng et~al.}(1998)\citenamefont{Feng, Moroi, Murayama,
  and Schnapka}}]{Feng:1997tn}
\bibinfo{author}{\bibfnamefont{J.~L.} \bibnamefont{Feng}},
  \bibinfo{author}{\bibfnamefont{T.}~\bibnamefont{Moroi}},
  \bibinfo{author}{\bibfnamefont{H.}~\bibnamefont{Murayama}}, \bibnamefont{and}
  \bibinfo{author}{\bibfnamefont{E.}~\bibnamefont{Schnapka}},
  \bibinfo{journal}{Phys. Rev. D} \textbf{\bibinfo{volume}{57}},
  \bibinfo{pages}{5875} (\bibinfo{year}{1998}), \eprint{hep-ph/9709411}.

\bibitem[{\citenamefont{Cornella et~al.}(2020)\citenamefont{Cornella, Paradisi,
  and Sumensari}}]{Cornella:2019uxs}
\bibinfo{author}{\bibfnamefont{C.}~\bibnamefont{Cornella}},
  \bibinfo{author}{\bibfnamefont{P.}~\bibnamefont{Paradisi}}, \bibnamefont{and}
  \bibinfo{author}{\bibfnamefont{O.}~\bibnamefont{Sumensari}},
  \bibinfo{journal}{JHEP} \textbf{\bibinfo{volume}{01}}, \bibinfo{pages}{158}
  (\bibinfo{year}{2020}), \eprint{1911.06279}.

\bibitem[{\citenamefont{Holdom}(1986)}]{Holdom:1985ag}
\bibinfo{author}{\bibfnamefont{B.}~\bibnamefont{Holdom}},
  \bibinfo{journal}{Phys. Lett. B} \textbf{\bibinfo{volume}{166}},
  \bibinfo{pages}{196} (\bibinfo{year}{1986}).

\bibitem[{\citenamefont{Fayet}(2017)}]{Fayet:2016nyc}
\bibinfo{author}{\bibfnamefont{P.}~\bibnamefont{Fayet}}, \bibinfo{journal}{Eur.
  Phys. J. C} \textbf{\bibinfo{volume}{77}}, \bibinfo{pages}{53}
  (\bibinfo{year}{2017}), \eprint{1611.05357}.

\bibitem[{\citenamefont{Altarelli et~al.}(1989)\citenamefont{Altarelli, Mele,
  and Ruiz-Altaba}}]{Altarelli:1989ff}
\bibinfo{author}{\bibfnamefont{G.}~\bibnamefont{Altarelli}},
  \bibinfo{author}{\bibfnamefont{B.}~\bibnamefont{Mele}}, \bibnamefont{and}
  \bibinfo{author}{\bibfnamefont{M.}~\bibnamefont{Ruiz-Altaba}},
  \bibinfo{journal}{Z. Phys. C} \textbf{\bibinfo{volume}{45}},
  \bibinfo{pages}{109} (\bibinfo{year}{1989}), \bibinfo{note}{[Erratum:
  Z.Phys.C 47, 676 (1990)]}.

\bibitem[{\citenamefont{Khachatryan et~al.}(2017)}]{CMS:2016ifc}
\bibinfo{author}{\bibfnamefont{V.}~\bibnamefont{Khachatryan}}
  \bibnamefont{et~al.} (\bibinfo{collaboration}{CMS Collaboration}),
  \bibinfo{journal}{Phys. Lett. B} \textbf{\bibinfo{volume}{770}},
  \bibinfo{pages}{278} (\bibinfo{year}{2017}), \eprint{1612.09274}.

\bibitem[{\citenamefont{Aaboud et~al.}(2017)}]{ATLAS:2017eqx}
\bibinfo{author}{\bibfnamefont{M.}~\bibnamefont{Aaboud}} \bibnamefont{et~al.}
  (\bibinfo{collaboration}{ATLAS Collaboration}), \bibinfo{journal}{Phys. Rev.
  D} \textbf{\bibinfo{volume}{96}}, \bibinfo{pages}{052004}
  (\bibinfo{year}{2017}), \eprint{1703.09127}.

\bibitem[{\citenamefont{Aad et~al.}(2019{\natexlab{a}})}]{ATLAS:2019erb}
\bibinfo{author}{\bibfnamefont{G.}~\bibnamefont{Aad}} \bibnamefont{et~al.}
  (\bibinfo{collaboration}{ATLAS Collaboration}), \bibinfo{journal}{Phys. Lett.
  B} \textbf{\bibinfo{volume}{796}}, \bibinfo{pages}{68}
  (\bibinfo{year}{2019}{\natexlab{a}}), \eprint{1903.06248}.

\bibitem[{\citenamefont{Aad et~al.}(2021)}]{ATLAS:2020uiq}
\bibinfo{author}{\bibfnamefont{G.}~\bibnamefont{Aad}} \bibnamefont{et~al.}
  (\bibinfo{collaboration}{ATLAS Collaboration}), \bibinfo{journal}{JHEP}
  \textbf{\bibinfo{volume}{02}}, \bibinfo{pages}{226} (\bibinfo{year}{2021}),
  \eprint{2011.05259}.

\bibitem[{\citenamefont{Banerjee et~al.}(2017)}]{NA64:2016oww}
\bibinfo{author}{\bibfnamefont{D.}~\bibnamefont{Banerjee}} \bibnamefont{et~al.}
  (\bibinfo{collaboration}{NA64 Collaboration}), \bibinfo{journal}{Phys. Rev.
  Lett.} \textbf{\bibinfo{volume}{118}}, \bibinfo{pages}{011802}
  (\bibinfo{year}{2017}), \eprint{1610.02988}.

\bibitem[{\citenamefont{Banerjee et~al.}(2018{\natexlab{a}})}]{NA64:2017vtt}
\bibinfo{author}{\bibfnamefont{D.}~\bibnamefont{Banerjee}} \bibnamefont{et~al.}
  (\bibinfo{collaboration}{NA64 Collaboration}), \bibinfo{journal}{Phys. Rev.
  D} \textbf{\bibinfo{volume}{97}}, \bibinfo{pages}{072002}
  (\bibinfo{year}{2018}{\natexlab{a}}), \eprint{1710.00971}.

\bibitem[{\citenamefont{Banerjee et~al.}(2018{\natexlab{b}})}]{NA64:2018lsq}
\bibinfo{author}{\bibfnamefont{D.}~\bibnamefont{Banerjee}} \bibnamefont{et~al.}
  (\bibinfo{collaboration}{NA64 Collaboration}), \bibinfo{journal}{Phys. Rev.
  Lett.} \textbf{\bibinfo{volume}{120}}, \bibinfo{pages}{231802}
  (\bibinfo{year}{2018}{\natexlab{b}}), \eprint{1803.07748}.

\bibitem[{\citenamefont{Cazzaniga et~al.}(2021)}]{NA64:2021acr}
\bibinfo{author}{\bibfnamefont{C.}~\bibnamefont{Cazzaniga}}
  \bibnamefont{et~al.} (\bibinfo{collaboration}{NA64 Collaboration}),
  \bibinfo{journal}{Eur. Phys. J. C} \textbf{\bibinfo{volume}{81}},
  \bibinfo{pages}{959} (\bibinfo{year}{2021}), \eprint{2107.02021}.

\bibitem[{\citenamefont{Andreev et~al.}(2022{\natexlab{a}})}]{NA64:2022rme}
\bibinfo{author}{\bibfnamefont{Y.~M.} \bibnamefont{Andreev}}
  \bibnamefont{et~al.} (\bibinfo{collaboration}{NA64 Collaboration}),
  \bibinfo{journal}{Phys. Rev. D} \textbf{\bibinfo{volume}{106}},
  \bibinfo{pages}{032015} (\bibinfo{year}{2022}{\natexlab{a}}),
  \eprint{2206.03101}.

\bibitem[{\citenamefont{Andreev et~al.}(2022{\natexlab{b}})}]{Andreev:2022hxz}
\bibinfo{author}{\bibfnamefont{Y.~M.} \bibnamefont{Andreev}}
  \bibnamefont{et~al.} (\bibinfo{collaboration}{NA64 Collaboration})
  (\bibinfo{year}{2022}{\natexlab{b}}), \eprint{2207.09979}.

\bibitem[{\citenamefont{Pankov et~al.}(2020)\citenamefont{Pankov, Osland,
  Serenkova, and Bednyakov}}]{Pankov:2019yzr}
\bibinfo{author}{\bibfnamefont{A.~A.} \bibnamefont{Pankov}},
  \bibinfo{author}{\bibfnamefont{P.}~\bibnamefont{Osland}},
  \bibinfo{author}{\bibfnamefont{I.~A.} \bibnamefont{Serenkova}},
  \bibnamefont{and} \bibinfo{author}{\bibfnamefont{V.~A.}
  \bibnamefont{Bednyakov}}, \bibinfo{journal}{Eur. Phys. J. C}
  \textbf{\bibinfo{volume}{80}}, \bibinfo{pages}{503} (\bibinfo{year}{2020}),
  \eprint{1912.02106}.

\bibitem[{\citenamefont{Osland et~al.}(2022)\citenamefont{Osland, Pankov, and
  Serenkova}}]{Osland:2022ryb}
\bibinfo{author}{\bibfnamefont{P.}~\bibnamefont{Osland}},
  \bibinfo{author}{\bibfnamefont{A.~A.} \bibnamefont{Pankov}},
  \bibnamefont{and} \bibinfo{author}{\bibfnamefont{I.~A.}
  \bibnamefont{Serenkova}} (\bibinfo{year}{2022}), \eprint{2206.01438}.

\bibitem[{\citenamefont{Osland et~al.}(2021)\citenamefont{Osland, Pankov, and
  Serenkova}}]{Osland:2020onj}
\bibinfo{author}{\bibfnamefont{P.}~\bibnamefont{Osland}},
  \bibinfo{author}{\bibfnamefont{A.~A.} \bibnamefont{Pankov}},
  \bibnamefont{and} \bibinfo{author}{\bibfnamefont{I.~A.}
  \bibnamefont{Serenkova}}, \bibinfo{journal}{Phys. Rev. D}
  \textbf{\bibinfo{volume}{103}}, \bibinfo{pages}{053009}
  (\bibinfo{year}{2021}), \eprint{2012.13930}.

\bibitem[{\citenamefont{Fuks and Ruiz}(2017)}]{Fuks:2017vtl}
\bibinfo{author}{\bibfnamefont{B.}~\bibnamefont{Fuks}} \bibnamefont{and}
  \bibinfo{author}{\bibfnamefont{R.}~\bibnamefont{Ruiz}},
  \bibinfo{journal}{JHEP} \textbf{\bibinfo{volume}{05}}, \bibinfo{pages}{032}
  (\bibinfo{year}{2017}), \eprint{1701.05263}.

\bibitem[{\citenamefont{Heeck}(2017)}]{Heeck:2016xwg}
\bibinfo{author}{\bibfnamefont{J.}~\bibnamefont{Heeck}},
  \bibinfo{journal}{Phys. Rev. D} \textbf{\bibinfo{volume}{95}},
  \bibinfo{pages}{015022} (\bibinfo{year}{2017}), \eprint{1610.07623}.

\bibitem[{\citenamefont{Belyaev et~al.}(2022)\citenamefont{Belyaev, Deandrea,
  Moretti, Panizzi, and Thongyoi}}]{Belyaev:2022shr}
\bibinfo{author}{\bibfnamefont{A.}~\bibnamefont{Belyaev}},
  \bibinfo{author}{\bibfnamefont{A.}~\bibnamefont{Deandrea}},
  \bibinfo{author}{\bibfnamefont{S.}~\bibnamefont{Moretti}},
  \bibinfo{author}{\bibfnamefont{L.}~\bibnamefont{Panizzi}}, \bibnamefont{and}
  \bibinfo{author}{\bibfnamefont{N.}~\bibnamefont{Thongyoi}}
  (\bibinfo{year}{2022}), \eprint{2204.03510}.

\bibitem[{\citenamefont{Eichten et~al.}(1984)\citenamefont{Eichten, Hinchliffe,
  Lane, and Quigg}}]{Eichten:1984eu}
\bibinfo{author}{\bibfnamefont{E.}~\bibnamefont{Eichten}},
  \bibinfo{author}{\bibfnamefont{I.}~\bibnamefont{Hinchliffe}},
  \bibinfo{author}{\bibfnamefont{K.~D.} \bibnamefont{Lane}}, \bibnamefont{and}
  \bibinfo{author}{\bibfnamefont{C.}~\bibnamefont{Quigg}},
  \bibinfo{journal}{Rev. Mod. Phys.} \textbf{\bibinfo{volume}{56}},
  \bibinfo{pages}{579} (\bibinfo{year}{1984}), \bibinfo{note}{[Addendum:
  Rev.Mod.Phys. 58, 1065--1073 (1986)]}.

\bibitem[{\citenamefont{Randall and
  Sundrum}(1999{\natexlab{a}})}]{Randall:1999ee}
\bibinfo{author}{\bibfnamefont{L.}~\bibnamefont{Randall}} \bibnamefont{and}
  \bibinfo{author}{\bibfnamefont{R.}~\bibnamefont{Sundrum}},
  \bibinfo{journal}{Phys. Rev. Lett.} \textbf{\bibinfo{volume}{83}},
  \bibinfo{pages}{3370} (\bibinfo{year}{1999}{\natexlab{a}}),
  \eprint{hep-ph/9905221}.

\bibitem[{\citenamefont{Randall and
  Sundrum}(1999{\natexlab{b}})}]{Randall:1999vf}
\bibinfo{author}{\bibfnamefont{L.}~\bibnamefont{Randall}} \bibnamefont{and}
  \bibinfo{author}{\bibfnamefont{R.}~\bibnamefont{Sundrum}},
  \bibinfo{journal}{Phys. Rev. Lett.} \textbf{\bibinfo{volume}{83}},
  \bibinfo{pages}{4690} (\bibinfo{year}{1999}{\natexlab{b}}),
  \eprint{hep-th/9906064}.

\bibitem[{\citenamefont{Davoudiasl et~al.}(2001)\citenamefont{Davoudiasl,
  Hewett, and Rizzo}}]{Davoudiasl:2000wi}
\bibinfo{author}{\bibfnamefont{H.}~\bibnamefont{Davoudiasl}},
  \bibinfo{author}{\bibfnamefont{J.~L.} \bibnamefont{Hewett}},
  \bibnamefont{and} \bibinfo{author}{\bibfnamefont{T.~G.} \bibnamefont{Rizzo}},
  \bibinfo{journal}{Phys. Rev. D} \textbf{\bibinfo{volume}{63}},
  \bibinfo{pages}{075004} (\bibinfo{year}{2001}), \eprint{hep-ph/0006041}.

\bibitem[{\citenamefont{Lane and Mrenna}(2003)}]{Lane:2002sm}
\bibinfo{author}{\bibfnamefont{K.}~\bibnamefont{Lane}} \bibnamefont{and}
  \bibinfo{author}{\bibfnamefont{S.}~\bibnamefont{Mrenna}},
  \bibinfo{journal}{Phys. Rev. D} \textbf{\bibinfo{volume}{67}},
  \bibinfo{pages}{115011} (\bibinfo{year}{2003}), \eprint{hep-ph/0210299}.

\bibitem[{\citenamefont{Eichten and Lane}(2008)}]{Eichten:2007sx}
\bibinfo{author}{\bibfnamefont{E.}~\bibnamefont{Eichten}} \bibnamefont{and}
  \bibinfo{author}{\bibfnamefont{K.}~\bibnamefont{Lane}},
  \bibinfo{journal}{Phys. Lett. B} \textbf{\bibinfo{volume}{669}},
  \bibinfo{pages}{235} (\bibinfo{year}{2008}), \eprint{0706.2339}.

\bibitem[{\citenamefont{Agashe et~al.}(2005)\citenamefont{Agashe, Contino, and
  Pomarol}}]{Agashe:2004rs}
\bibinfo{author}{\bibfnamefont{K.}~\bibnamefont{Agashe}},
  \bibinfo{author}{\bibfnamefont{R.}~\bibnamefont{Contino}}, \bibnamefont{and}
  \bibinfo{author}{\bibfnamefont{A.}~\bibnamefont{Pomarol}},
  \bibinfo{journal}{Nucl. Phys. B} \textbf{\bibinfo{volume}{719}},
  \bibinfo{pages}{165} (\bibinfo{year}{2005}), \eprint{hep-ph/0412089}.

\bibitem[{\citenamefont{Giudice et~al.}(2007)\citenamefont{Giudice, Grojean,
  Pomarol, and Rattazzi}}]{Giudice:2007fh}
\bibinfo{author}{\bibfnamefont{G.~F.} \bibnamefont{Giudice}},
  \bibinfo{author}{\bibfnamefont{C.}~\bibnamefont{Grojean}},
  \bibinfo{author}{\bibfnamefont{A.}~\bibnamefont{Pomarol}}, \bibnamefont{and}
  \bibinfo{author}{\bibfnamefont{R.}~\bibnamefont{Rattazzi}},
  \bibinfo{journal}{JHEP} \textbf{\bibinfo{volume}{06}}, \bibinfo{pages}{045}
  (\bibinfo{year}{2007}), \eprint{hep-ph/0703164}.

\bibitem[{\citenamefont{Aaboud et~al.}(2019)}]{ATLAS:2019isd}
\bibinfo{author}{\bibfnamefont{M.}~\bibnamefont{Aaboud}} \bibnamefont{et~al.}
  (\bibinfo{collaboration}{ATLAS}), \bibinfo{journal}{Phys. Lett. B}
  \textbf{\bibinfo{volume}{798}}, \bibinfo{pages}{134942}
  (\bibinfo{year}{2019}), \eprint{1904.12679}.

\bibitem[{\citenamefont{Aad et~al.}(2019{\natexlab{b}})}]{ATLAS:2019lsy}
\bibinfo{author}{\bibfnamefont{G.}~\bibnamefont{Aad}} \bibnamefont{et~al.}
  (\bibinfo{collaboration}{ATLAS}), \bibinfo{journal}{Phys. Rev. D}
  \textbf{\bibinfo{volume}{100}}, \bibinfo{pages}{052013}
  (\bibinfo{year}{2019}{\natexlab{b}}), \eprint{1906.05609}.

\bibitem[{\citenamefont{Sirunyan et~al.}(2019)}]{CMS:2018iye}
\bibinfo{author}{\bibfnamefont{A.~M.} \bibnamefont{Sirunyan}}
  \bibnamefont{et~al.} (\bibinfo{collaboration}{CMS}), \bibinfo{journal}{JHEP}
  \textbf{\bibinfo{volume}{03}}, \bibinfo{pages}{170} (\bibinfo{year}{2019}),
  \eprint{1811.00806}.

\bibitem[{\citenamefont{Tumasyan et~al.}(2022)}]{CMS:2022yjm}
\bibinfo{author}{\bibfnamefont{A.}~\bibnamefont{Tumasyan}} \bibnamefont{et~al.}
  (\bibinfo{collaboration}{CMS}), \bibinfo{journal}{JHEP}
  \textbf{\bibinfo{volume}{07}}, \bibinfo{pages}{067} (\bibinfo{year}{2022}),
  \eprint{2202.06075}.

\bibitem[{\citenamefont{Stueckelberg}(1938)}]{Stueckelberg:1938zz}
\bibinfo{author}{\bibfnamefont{E.~C.~G.} \bibnamefont{Stueckelberg}},
  \bibinfo{journal}{Helv. Phys. Acta} \textbf{\bibinfo{volume}{11}},
  \bibinfo{pages}{299} (\bibinfo{year}{1938}).

\bibitem[{\citenamefont{Bell et~al.}(2017)\citenamefont{Bell, Cai, and
  Leane}}]{Bell:2016uhg}
\bibinfo{author}{\bibfnamefont{N.~F.} \bibnamefont{Bell}},
  \bibinfo{author}{\bibfnamefont{Y.}~\bibnamefont{Cai}}, \bibnamefont{and}
  \bibinfo{author}{\bibfnamefont{R.~K.} \bibnamefont{Leane}},
  \bibinfo{journal}{JCAP} \textbf{\bibinfo{volume}{01}}, \bibinfo{pages}{039}
  (\bibinfo{year}{2017}), \eprint{1610.03063}.

\bibitem[{\citenamefont{Gherghetta et~al.}(2019)\citenamefont{Gherghetta,
  Kersten, Olive, and Pospelov}}]{Gherghetta:2019coi}
\bibinfo{author}{\bibfnamefont{T.}~\bibnamefont{Gherghetta}},
  \bibinfo{author}{\bibfnamefont{J.}~\bibnamefont{Kersten}},
  \bibinfo{author}{\bibfnamefont{K.}~\bibnamefont{Olive}}, \bibnamefont{and}
  \bibinfo{author}{\bibfnamefont{M.}~\bibnamefont{Pospelov}},
  \bibinfo{journal}{Phys. Rev. D} \textbf{\bibinfo{volume}{100}},
  \bibinfo{pages}{095001} (\bibinfo{year}{2019}), \eprint{1909.00696}.

\bibitem[{\citenamefont{Delbourgo and Thompson}(1986)}]{Delbourgo:1986wz}
\bibinfo{author}{\bibfnamefont{R.}~\bibnamefont{Delbourgo}} \bibnamefont{and}
  \bibinfo{author}{\bibfnamefont{G.}~\bibnamefont{Thompson}},
  \bibinfo{journal}{Phys. Rev. Lett.} \textbf{\bibinfo{volume}{57}},
  \bibinfo{pages}{2610} (\bibinfo{year}{1986}).

\bibitem[{\citenamefont{Glashow et~al.}(1970)\citenamefont{Glashow, Iliopoulos,
  and Maiani}}]{GIM}
\bibinfo{author}{\bibfnamefont{S.~L.} \bibnamefont{Glashow}},
  \bibinfo{author}{\bibfnamefont{J.}~\bibnamefont{Iliopoulos}},
  \bibnamefont{and} \bibinfo{author}{\bibfnamefont{L.}~\bibnamefont{Maiani}},
  \bibinfo{journal}{Phys. Rev. D} \textbf{\bibinfo{volume}{2}},
  \bibinfo{pages}{1285} (\bibinfo{year}{1970}).

\bibitem[{\citenamefont{Kors and Nath}(2005)}]{Kors:2005uz}
\bibinfo{author}{\bibfnamefont{B.}~\bibnamefont{Kors}} \bibnamefont{and}
  \bibinfo{author}{\bibfnamefont{P.}~\bibnamefont{Nath}},
  \bibinfo{journal}{JHEP} \textbf{\bibinfo{volume}{07}}, \bibinfo{pages}{069}
  (\bibinfo{year}{2005}), \eprint{hep-ph/0503208}.

\bibitem[{\citenamefont{Webber et~al.}(2011)}]{MuLan:2010shf}
\bibinfo{author}{\bibfnamefont{D.~M.} \bibnamefont{Webber}}
  \bibnamefont{et~al.} (\bibinfo{collaboration}{MuLan Collaboration}),
  \bibinfo{journal}{Phys. Rev. Lett.} \textbf{\bibinfo{volume}{106}},
  \bibinfo{pages}{041803} (\bibinfo{year}{2011}), \eprint{1010.0991}.

\bibitem[{\citenamefont{Tishchenko et~al.}(2013)}]{MuLan:2012sih}
\bibinfo{author}{\bibfnamefont{V.}~\bibnamefont{Tishchenko}}
  \bibnamefont{et~al.} (\bibinfo{collaboration}{MuLan Collaboration}),
  \bibinfo{journal}{Phys. Rev. D} \textbf{\bibinfo{volume}{87}},
  \bibinfo{pages}{052003} (\bibinfo{year}{2013}), \eprint{1211.0960}.

\bibitem[{\citenamefont{Gorringe and Hertzog}(2015)}]{Gorringe:2015cma}
\bibinfo{author}{\bibfnamefont{T.~P.} \bibnamefont{Gorringe}} \bibnamefont{and}
  \bibinfo{author}{\bibfnamefont{D.~W.} \bibnamefont{Hertzog}},
  \bibinfo{journal}{Prog. Part. Nucl. Phys.} \textbf{\bibinfo{volume}{84}},
  \bibinfo{pages}{73} (\bibinfo{year}{2015}), \eprint{1506.01465}.

\bibitem[{\citenamefont{Crivellin et~al.}(2021)\citenamefont{Crivellin,
  Hoferichter, and Manzari}}]{Crivellin:2021njn}
\bibinfo{author}{\bibfnamefont{A.}~\bibnamefont{Crivellin}},
  \bibinfo{author}{\bibfnamefont{M.}~\bibnamefont{Hoferichter}},
  \bibnamefont{and} \bibinfo{author}{\bibfnamefont{C.~A.}
  \bibnamefont{Manzari}}, \bibinfo{journal}{Phys. Rev. Lett.}
  \textbf{\bibinfo{volume}{127}}, \bibinfo{pages}{071801}
  (\bibinfo{year}{2021}), \eprint{2102.02825}.

\bibitem[{\citenamefont{Zyla et~al.}(2020)}]{ParticleDataGroup:2020ssz}
\bibinfo{author}{\bibfnamefont{P.~A.} \bibnamefont{Zyla}} \bibnamefont{et~al.}
  (\bibinfo{collaboration}{Particle Data Group}), \bibinfo{journal}{PTEP}
  \textbf{\bibinfo{volume}{2020}}, \bibinfo{pages}{083C01}
  (\bibinfo{year}{2020}).

\bibitem[{ATL(2023)}]{ATLAS:2023vxg}
 (\bibinfo{year}{2023}), \eprint{2301.09342}.

\bibitem[{\citenamefont{Lindner et~al.}(2018)\citenamefont{Lindner, Platscher,
  and Queiroz}}]{Lindner:2016bgg}
\bibinfo{author}{\bibfnamefont{M.}~\bibnamefont{Lindner}},
  \bibinfo{author}{\bibfnamefont{M.}~\bibnamefont{Platscher}},
  \bibnamefont{and} \bibinfo{author}{\bibfnamefont{F.~S.}
  \bibnamefont{Queiroz}}, \bibinfo{journal}{Phys. Rept.}
  \textbf{\bibinfo{volume}{731}}, \bibinfo{pages}{1} (\bibinfo{year}{2018}),
  \eprint{1610.06587}.

\bibitem[{\citenamefont{Langacker}(2009)}]{Langacker:2008yv}
\bibinfo{author}{\bibfnamefont{P.}~\bibnamefont{Langacker}},
  \bibinfo{journal}{Rev. Mod. Phys.} \textbf{\bibinfo{volume}{81}},
  \bibinfo{pages}{1199} (\bibinfo{year}{2009}), \eprint{0801.1345}.

\bibitem[{\citenamefont{Freedman et~al.}(1993)}]{Freedman:1993kz}
\bibinfo{author}{\bibfnamefont{S.~J.} \bibnamefont{Freedman}}
  \bibnamefont{et~al.}, \bibinfo{journal}{Phys. Rev. D}
  \textbf{\bibinfo{volume}{47}}, \bibinfo{pages}{811} (\bibinfo{year}{1993}).

\bibitem[{\citenamefont{Andreev et~al.}(2022{\natexlab{c}})}]{NA64:2022yly}
\bibinfo{author}{\bibfnamefont{Y.~M.} \bibnamefont{Andreev}}
  \bibnamefont{et~al.} (\bibinfo{collaboration}{NA64}), \bibinfo{journal}{Phys.
  Rev. Lett.} \textbf{\bibinfo{volume}{129}}, \bibinfo{pages}{161801}
  (\bibinfo{year}{2022}{\natexlab{c}}), \eprint{2207.09979}.

\bibitem[{\citenamefont{Depero et~al.}(2020)}]{NA64:2020xxh}
\bibinfo{author}{\bibfnamefont{E.}~\bibnamefont{Depero}} \bibnamefont{et~al.}
  (\bibinfo{collaboration}{NA64}), \bibinfo{journal}{Eur. Phys. J. C}
  \textbf{\bibinfo{volume}{80}}, \bibinfo{pages}{1159} (\bibinfo{year}{2020}),
  \eprint{2009.02756}.

\bibitem[{\citenamefont{Banerjee et~al.}(2020)}]{NA64:2019auh}
\bibinfo{author}{\bibfnamefont{D.}~\bibnamefont{Banerjee}} \bibnamefont{et~al.}
  (\bibinfo{collaboration}{NA64}), \bibinfo{journal}{Phys. Rev. D}
  \textbf{\bibinfo{volume}{101}}, \bibinfo{pages}{071101}
  (\bibinfo{year}{2020}), \eprint{1912.11389}.

\bibitem[{\citenamefont{Banerjee et~al.}(2019)}]{Banerjee:2019pds}
\bibinfo{author}{\bibfnamefont{D.}~\bibnamefont{Banerjee}}
  \bibnamefont{et~al.}, \bibinfo{journal}{Phys. Rev. Lett.}
  \textbf{\bibinfo{volume}{123}}, \bibinfo{pages}{121801}
  (\bibinfo{year}{2019}), \eprint{1906.00176}.

\bibitem[{\citenamefont{Kahn et~al.}(2018)\citenamefont{Kahn, Krnjaic, Tran,
  and Whitbeck}}]{Kahn:2018cqs}
\bibinfo{author}{\bibfnamefont{Y.}~\bibnamefont{Kahn}},
  \bibinfo{author}{\bibfnamefont{G.}~\bibnamefont{Krnjaic}},
  \bibinfo{author}{\bibfnamefont{N.}~\bibnamefont{Tran}}, \bibnamefont{and}
  \bibinfo{author}{\bibfnamefont{A.}~\bibnamefont{Whitbeck}},
  \bibinfo{journal}{JHEP} \textbf{\bibinfo{volume}{09}}, \bibinfo{pages}{153}
  (\bibinfo{year}{2018}), \eprint{1804.03144}.

\bibitem[{\citenamefont{Mans}(2017)}]{Mans:2017vej}
\bibinfo{author}{\bibfnamefont{J.}~\bibnamefont{Mans}}
  (\bibinfo{collaboration}{LDMX}), \bibinfo{journal}{EPJ Web Conf.}
  \textbf{\bibinfo{volume}{142}}, \bibinfo{pages}{01020}
  (\bibinfo{year}{2017}).

\bibitem[{\citenamefont{Profumo and Queiroz}(2014)}]{Profumo:2013sca}
\bibinfo{author}{\bibfnamefont{S.}~\bibnamefont{Profumo}} \bibnamefont{and}
  \bibinfo{author}{\bibfnamefont{F.~S.} \bibnamefont{Queiroz}},
  \bibinfo{journal}{Eur. Phys. J. C} \textbf{\bibinfo{volume}{74}},
  \bibinfo{pages}{2960} (\bibinfo{year}{2014}), \eprint{1307.7802}.

\bibitem[{\citenamefont{Kelso et~al.}(2014)\citenamefont{Kelso, Pinheiro,
  Queiroz, and Shepherd}}]{Kelso:2013zfa}
\bibinfo{author}{\bibfnamefont{C.}~\bibnamefont{Kelso}},
  \bibinfo{author}{\bibfnamefont{P.~R.~D.} \bibnamefont{Pinheiro}},
  \bibinfo{author}{\bibfnamefont{F.~S.} \bibnamefont{Queiroz}},
  \bibnamefont{and} \bibinfo{author}{\bibfnamefont{W.}~\bibnamefont{Shepherd}},
  \bibinfo{journal}{Eur. Phys. J. C} \textbf{\bibinfo{volume}{74}},
  \bibinfo{pages}{2808} (\bibinfo{year}{2014}), \eprint{1312.0051}.

\bibitem[{\citenamefont{Alves et~al.}(2015)\citenamefont{Alves, Berlin,
  Profumo, and Queiroz}}]{Alves:2015pea}
\bibinfo{author}{\bibfnamefont{A.}~\bibnamefont{Alves}},
  \bibinfo{author}{\bibfnamefont{A.}~\bibnamefont{Berlin}},
  \bibinfo{author}{\bibfnamefont{S.}~\bibnamefont{Profumo}}, \bibnamefont{and}
  \bibinfo{author}{\bibfnamefont{F.~S.} \bibnamefont{Queiroz}},
  \bibinfo{journal}{Phys. Rev. D} \textbf{\bibinfo{volume}{92}},
  \bibinfo{pages}{083004} (\bibinfo{year}{2015}), \eprint{1501.03490}.

\bibitem[{\citenamefont{Patel}(2015)}]{Patel:2015tea}
\bibinfo{author}{\bibfnamefont{H.~H.} \bibnamefont{Patel}},
  \bibinfo{journal}{Comput. Phys. Commun.} \textbf{\bibinfo{volume}{197}},
  \bibinfo{pages}{276} (\bibinfo{year}{2015}), \eprint{1503.01469}.

\bibitem[{\citenamefont{Shtabovenko}(2017)}]{Shtabovenko:2016whf}
\bibinfo{author}{\bibfnamefont{V.}~\bibnamefont{Shtabovenko}},
  \bibinfo{journal}{Comput. Phys. Commun.} \textbf{\bibinfo{volume}{218}},
  \bibinfo{pages}{48} (\bibinfo{year}{2017}), \eprint{1611.06793}.

\bibitem[{\citenamefont{Shtabovenko et~al.}(2020)\citenamefont{Shtabovenko,
  Mertig, and Orellana}}]{Shtabovenko:2020gxv}
\bibinfo{author}{\bibfnamefont{V.}~\bibnamefont{Shtabovenko}},
  \bibinfo{author}{\bibfnamefont{R.}~\bibnamefont{Mertig}}, \bibnamefont{and}
  \bibinfo{author}{\bibfnamefont{F.}~\bibnamefont{Orellana}},
  \bibinfo{journal}{Comput. Phys. Commun.} \textbf{\bibinfo{volume}{256}},
  \bibinfo{pages}{107478} (\bibinfo{year}{2020}), \eprint{2001.04407}.

\bibitem[{\citenamefont{Shtabovenko et~al.}(2016)\citenamefont{Shtabovenko,
  Mertig, and Orellana}}]{Shtabovenko:2016sxi}
\bibinfo{author}{\bibfnamefont{V.}~\bibnamefont{Shtabovenko}},
  \bibinfo{author}{\bibfnamefont{R.}~\bibnamefont{Mertig}}, \bibnamefont{and}
  \bibinfo{author}{\bibfnamefont{F.}~\bibnamefont{Orellana}},
  \bibinfo{journal}{Comput. Phys. Commun.} \textbf{\bibinfo{volume}{207}},
  \bibinfo{pages}{432} (\bibinfo{year}{2016}), \eprint{1601.01167}.

\bibitem[{\citenamefont{Mertig et~al.}(1991)\citenamefont{Mertig, Bohm, and
  Denner}}]{Mertig:1990an}
\bibinfo{author}{\bibfnamefont{R.}~\bibnamefont{Mertig}},
  \bibinfo{author}{\bibfnamefont{M.}~\bibnamefont{Bohm}}, \bibnamefont{and}
  \bibinfo{author}{\bibfnamefont{A.}~\bibnamefont{Denner}},
  \bibinfo{journal}{Comput. Phys. Commun.} \textbf{\bibinfo{volume}{64}},
  \bibinfo{pages}{345} (\bibinfo{year}{1991}).

\end{thebibliography}

\end{document}